\documentclass[aps,prb,groupedaddress,showpacs,floats,twocolumn,floatfix]{revtex4}
\usepackage{graphicx}
\usepackage{amsmath}
\usepackage{dcolumn}

\begin{document}



\title{Second-order electronic correlation effects 
in a one-dimensional metal}



\author{Rafa{\l} Podeszwa}


\author{Leszek Z. Stolarczyk}


\affiliation{Department of Chemistry, University
of Warsaw, Pasteura 1, PL-02-093 Warsaw, Poland}


\date{\today}

\begin{abstract}

The Pariser-Parr-Pople (PPP) model of a single-band one-dimensional (1D) metal is studied at the Hartree-Fock  level,  and by using the second-order perturbation theory of the electronic correlation. The PPP model provides an extension of the Hubbard model by properly accounting for the long-range character of the electron-electron repulsion. Both finite and infinite version of the 1D-metal model are considered within the PPP and Hubbard approximations. Calculated are the second-order electronic-correlation corrections to the total energy, and to the electronic-energy bands. Our results for the PPP model of 1D metal show qualitative similarity to the coupled-cluster results for the 3D electron-gas model. The picture of the 1D-metal model that emerges from the present study provides a support for the hypothesis that the normal metallic state of the 1D metal is different from the ground state.

\end{abstract}



\maketitle



\section{\label{sec:intro}%
Introduction}

  The quantum-mechanical description of a metallic system encounters difficulties due to the Fermi-level degeneracy and the long-range character of the Coulombic
interactions.  These difficulties seem to be resolved in the case of the three-dimensional (3D) electron gas,~\cite{hedin:69} which serves as the primary model of
isotropic metallic systems. On the other hand, the highly anisotropic metallic systems, exhibiting quasi-2D or quasi-1D metallic behavior, are much worse understood. For such systems, it is natural to use lattice models. However, the 2D- or 1D-lattice models are usually studied within the Hubbard approximation,~\cite{hubbard:63} in which the Coulombic interactions are completely screened outside of a given atomic (or molecular) site. A very nice feature of this approximation is the availability of exact solutions for several systems, including the ground state of the simple 1D lattice.~\cite{lieb:68} A progress in the experimental investigations of the so-called quantum wires~\cite{goni:91} has inspired 
recent theoretical studies of the 1D metals which use more realistic models, stressing the importance of the long-range Coulombic
interactions among electrons.~\cite{schulz:93,poilblanc:97,hatsugai:97,fano:99,capponi:00,wang:01,lee:02,valenzuela:03} 
In the present paper we consider a simple model of the 1D metal in which
no {\it a priori\/} screening of the interelectronic interactions is assumed. By using the M{\o}ller-Plesset (MP) perturbation
theory,~\cite{moller:34,brueckner:55,goldstone:57} known also as the many-body perturbation theory (MBPT),\cite{bartlett:81} we study the second-order electronic-correlation corrections to
the total energy and to the electronic-energy band spectrum of this model. A comparison is made with the Hubbard model, to uncover qualitative differences between the
two approaches.

   The 1D-lattice model of a metallic system is an idealization of such quasi-1D metals as conducting polymers [polyacetylene, (SN)$_x$, chains of chelated
transition-metal complexes] and certain classes of segregated-stack donor-acceptor molecular crystals (TTF-TCNQ, Bechgaard salts, etc.); the respective references may be
found in the review article by Bryce and Murphy.~\cite{bryce:84} Such a model may be viewed as describing a single chain of subsystems (atoms, molecular fragments,
molecules), exhibiting a translational periodicity in one dimension (a 3D character of the subsystems is implicitly assumed). The simplest single-band 1D-metal model with
the Coulombic interactions may be formulated on the basis of the Pariser-Parr-Pople (PPP) model~\cite{pariser:53a,pariser:53b,pople:53} of $\pi$-electron molecules.

However, despite of being a direct extension of the Hubbard model, the PPP model is virtually unknown to the condensed-matter theorists (recent papers by Fano {\it et
al.\/}~\cite{fano:98,fano:99} are noteworthy exceptions). In Sec.~\ref{sec:PPP} we give some details of the 1D-metal model based on the PPP theory.
It is known that the long-range Coulomb interactions in the extended-system limit result in conditionally convergent lattice sums, which have to be handled with
care. In the Hartree-Fock (HF) theory of polymers, a very efficient multipole-expansion technique was introduced by Piela and Delhalle.~\cite{piela:78,delhalle:80} For
small-energy-gap and metallic polymers, a special treatment of extremely slowly converging HF exchange-energy lattice sums becomes essential, see
Refs.~\onlinecite{stolarczyk:88} and~\onlinecite{jeziorska:90}, and references cited therein. Thus, the 1D-metal model of Sec.~\ref{sec:PPP} is highly nontrivial already
at the HF level, exhibiting the pathologies of the exact HF exchange which are characteristic to metallic systems of arbitrary dimensionality.~\cite{monkhorst:79} A
complete HF description of this model, for various band fillings, is given in Sec.~\ref{sec:HF}.

  The Hartree-Fock model of electronic systems provides a reference for a systematic inclusion of the electronic-correlation effects by the perturbation theory or the related
coupled-cluster (CC) method,~\cite{coester:58,cizek:66} up to the level described by the full configuration-interaction (FCI) 
method, which provides the exact solution in the algebraic approximation. 
 Both the perturbation and coupled-cluster methods are
nonvariational, but ensure the extensivity of the calculated correlation contributions, the property which is crucial in applications to extended systems. On the other
hand, the variational procedures employing a truncated configuration-interaction scheme lack the extensivity property, and are thus useless for extended systems. Within the last two decades, a
progress has been made in the electronic-correlation studies of 1D extended systems, by the
perturbation theory,~\cite{suhai:82,suhai:83a,suhai:83b,suhai:83c,suhai:92,suhai:93,suhai:94a,suhai:94b,suhai:95a,suhai:95b,sun:96a,sun:96b,sun:97,sun:98,sun:99,hirata:01} as well as the
coupled-cluster method.~\cite{hirata:01,forner:92,ye:93,knab:96,knab:97,forner:97} In comparison to molecular applications, these approaches to extended systems require special
techniques for handling the translational symmetry and infinite lattice sums. In metallic systems, one may expect additional problems due to the Fermi-level degeneracy
(the vanishing of the orbital-energy denominators).  Moreover, the severe numerical problems caused by the slow convergence of the Hartree-Fock exchange-energy terms are likely to
propagate into the electronic-correlation calculations, since the exchange and correlation contributions have to partially cancel each other, to effect the removal of the
HF-exchange pathologies. Some insight into these problems is provided by the performance of the perturbation and CC methods for the 3D electron gas.  It is known
that the perturbation theory  fails in this case, by predicting infinite values of the individual energy corrections.~\cite{gell-mann:57a,gell-mann:57b, handler:88} However,
meaningful results can be obtained by selectively summing up certain (infinite) classes of the energy contributions, see Gell-Mann and
Brueckner.~\cite{gell-mann:57a,gell-mann:57b} The coupled-cluster method provides another way of performing infinite-order summation of the perturbation-theory diagrams, and it proved to be very
successful in application to the 3D electron gas,~\cite{freeman:77,bishop:78,bishop:82,emrich:84} giving the electronic-correlation energies in a very good agreement with
the exact results obtained numerically by means of the Monte Carlo method.~\cite{ceperley:80}

  Unfortunately, the coupled-cluster method is expected to fail when applied to the ground state of the 1D metal. This negative conclusion derives from the extensive CC studies of the
PPP model of small cyclic polyenes (C$_M$H$_M$, $M = 6, 10, 14, 18, \ldots$) of $D_{M{\rm h}}$ symmetry, originated by Paldus and
coworkers~\cite{paldus:82,takahashi:83,paldus:84a,paldus:84b,paldus:84c,takahashi:85,piecuch:90a,piecuch:90b,piecuch:90c,piecuch:91,paldus:92,piecuch:92} (with some recent
complements by the present authors~\cite{podeszwa:02a,podeszwa:03,podeszwa:02c,podeszwa:02d}). In the PPP model of cyclic polyenes (annulenes), the strength of the electronic
correlation may be increased by (i) decreasing $|\beta|$ for a fixed $M$, and/or (ii) by increasing $M$ for a fixed $\beta$ [parameter $\beta (<0)$ represents the
so-called resonance integral of the PPP model]. By studying both routes, (i) and (ii), it was found~\cite{paldus:84a,paldus:84b} that the basic coupled-cluster
method, called CCD
(where D stands for the {\em connected\/} double excitations), breaks down in the strongly correlated regime of the model. 
In the CCD method one solves a set of quadratic equations (the CCD equations)
which supply the so-called $t_2$ amplitudes corresponding to the D excitations; the electronic-correlation energy is a (linear) function of these $t_2$ amplitudes. The
breakdown of the CCD method for annulenes was attributed to the increasing importance of the connected quadruple (Q) excitations, which, together with the connected triple
(T) excitations, contribute to the {\em exact\/} equations for the $t_2$ amplitudes through some coupling terms (in the CCD method these terms are simply neglected).
Paldus {\it et al.\/}~\cite{paldus:84a,paldus:84b} then showed that the coupling terms corresponding to Q excitations approximately cancel certain quadratic terms in the
CCD equations. This opened ways of improving the CCD method by some modification of the quadratic terms in the CCD equations, without the need of explicitly calculating
the Q contributions. The approximate coupled-pair theory, called ACP-D45 (or ACP, in short), introduced earlier,~\cite{jankowski:80} and the approximate coupled-pair theory
with quadruples (ACPQ), devised by Paldus {\it et al.},~\cite{paldus:84a,paldus:84b} are variants of the CCD method which incorporate the above idea. These methods proved
to be very effective for small cyclic polyenes, being convergent and giving the correlation energies close to the FCI values. However, recent calculations by Podeszwa
{\it et al.},~\cite{podeszwa:02a} showed that even the inclusion of the T and Q excitations in the CC operator (within the full CCSDTQ method~\cite{kucharski:92}) is
insufficient for getting the converged CC results in the strongly correlated regime, thus indicating that still higher connected excitations are necessary for a
proper description of these systems. Moreover, it was found that in the strongly correlated regime the $t_2$ amplitudes corresponding to the ACP and ACPQ methods deviate
markedly from those derived from the full configuration interaction  calculations, despite a good agreement between the corresponding correlation energies. Our recent application of the ACP and ACPQ
methods to large cyclic polyenes~\cite{podeszwa:02c} brought in negative results: no convergence was found for the ACPQ method for $M \ge 198$, and for the ACP method for
$M \ge 446$. Thus, it appears that no currently available CC method is capable of studying the 1D-metal limit ($M \rightarrow \infty)$ of cyclic polyenes.

  Recently, the density-matrix renormalization-group (DMRG) technique~\cite{white:92} emerges as a new promising tool for studying electronic systems.~\cite{chan:02}
Applied to the annulenes described within the PPP model ($M = 6$--$34$),~\cite{fano:98} the DMRG method was shown to approach the FCI accuracy for the ground-state energy.
By extrapolating the DMRG results obtained for finite rings (with $M$ up to $80$) to the extended-system limit, Fano {\it et al.\/}~\cite{fano:99} investigated the
asymptotic behavior of the elementary excitations, the spin and charge correlation functions, and the momentum distribution, finding agreement with the previous
bosonisation study by Schulz.~\cite{schulz:93} Unfortunately, no DMRG-based band-structure theory is available as yet.

 It seems that the electronic-correlation problem in the 1D metal poses quite a challenge for the available quantum many-body techniques. Nevertheless, the pioneering {\it ab initio\/} studies of the electronic-correlation effects in polymers by Suhai involved also 1D metallic systems: a chain of equidistant hydrogen
atoms,~\cite{suhai:82,suhai:83c,suhai:94a} (H)$_{\infty}$ (see also Ref.~\onlinecite{liegener:85}), the equidistant zig-zag form of polyacetylene~\cite{suhai:83b,suhai:83c,suhai:92,suhai:95a}, (CH)$_{\infty}$, and the equidistant zig-zag form of polysilene,~\cite{suhai:93,suhai:95b} (SiH)$_{\infty}$.
For these systems Suhai calculated the MP2,~\cite{suhai:82,suhai:83b,suhai:83c,suhai:92} MP3,~\cite{suhai:93} and MP4~\cite{suhai:94a,suhai:95a,suhai:95b}
correlation-energy contributions to the total energy per unit cell, where MP$n$ denotes the $n$-th order of the M{\o}ller-Plesset perturbation theory. These calculations provide a numerical evidence that, unlike for the 3D electron-gas model,~\cite{gell-mann:57a} for the 1D metals the vanishing denominators do not cause the divergence of the individual MP$n$ contributions. 

The present paper is organized as follows: \\
The Pariser-Parr-Pople model of a single-band 1D metal is described in Sec.~\ref{sec:PPP}. The Hartree-Fock results for this model are presented in Sec.~\ref{sec:HF}. The second-order electronic-correlation effects in the 1D metal are investigated in Sec.~\ref{sec:E2tot} (the correlation corrections to the total energy per unit cell) and in Sec.~\ref{sec:E2band} (the correlation corrections to the electronic-energy bands). The second-order corrections are calculated for finite systems and in the extended-system limit, for various band fillings. We point to some striking analogies between our perturbative results for the PPP 1D-metal model, and the coupled-cluster results for the 3D electron-gas model obtained by Bishop and L\"{u}hrmann,~\cite{bishop:78,bishop:82} and Emrich and Zabolitzky.~\cite{emrich:84} In the last section, Sec.~\ref{sec:disc}, we confront the picture of the PPP model of 1D metal that emerges from the present study with that of the DMRG calculations by Fano {\em et al.\/}\cite{fano:99} These two pictures can be reconciled by adopting a hypothesis that the normal metallic state of the 1D metal is different from the ground state.

\section{\label{sec:PPP}%
Pariser-Parr-Pople model of 1D metal}

  The Pariser-Parr-Pople (PPP) model~\cite{pariser:53a,pariser:53b,pople:53} was created for describing the electronic states of $\pi$-electron molecules. A built-in feature of this model is a qualitatively correct treatment of the electrostatic interactions [electron-electron, electron-(atomic core), and (atomic core)-(atomic core)] within a molecule. Below we reformulate the PPP model for the linear polyacetylene (see Ref.~\onlinecite{stolarczyk:88}) to provide a description of a more general 1D metal.  We consider a system composed of identical subsystems forming a regular chain, which is represented by a simple 1D lattice with some translational parameter $R^0$. We focus on the electronic-structure problem, neglecting the electron-phonon coupling, and keeping the geometry of our chain fixed. Subsystems may be atoms or molecular fragments, connected by chemical bonds (as in polymers), or molecules interacting via the Van der Waals forces. Although it is not essential for the considerations of the present paper, one may assume that the chain is embedded in some 3D crystalline matrix providing the chain with a rigid structure and supplying a 3D phonon spectrum. It is supposed that the number of electrons in the chain may be varied (e.g., due to the interactions with some doping agents) and an electron transfer between the neighboring subsystems is possible, leading eventually to a metallic 1D band structure. Such systems, with finite (but very large) number of subsystems, may work as wires in the molecular electronic devices of the future.~\cite{joachim:00}

\begin{figure}
\includegraphics[width=1.0\linewidth]{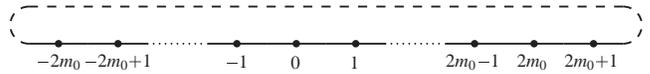}%
\caption{\label{fig:cyclic}
Periodic structure of the X$_M$ system, $M = 4m_0 + 2$.}
\end{figure}

  Let X$_M$ be a regular chain of subsystems X, hereafter called molecules, with the nearest-neighbor distance $R^0$. In order to arrive at the limit of an infinite chain
corresponding to the 1D metal, one has to build a finite cyclic model which leads to the fastest convergence for $M \rightarrow \infty$. Such a model corresponds to a
locally linear chain with periodic boundary conditions, which is depicted in Fig.~\ref{fig:cyclic}. For convenience, we choose $M = 4m_0 + 2$, $m_0 = 1,2, \ldots \,,$ as in
cyclic polyenes studied in Ref.~\onlinecite{podeszwa:02a}. The electronic structure of molecule X is assumed to be
frozen, except for the occupation number of some {\em outer\/} molecular orbital (MO) which may be varied from $0$ to $2$. In our PPP model, these (outer) MOs of all the molecules
in the X$_M$ system provide a basis set used in the quantum description of $N$ mobile electrons, $0 \leq N \leq 2M$. The ratio $N/M$ is equal to the mean occupation number
of the outer MO. The ground state of the chain may be of the closed-shell type only if the number of the mobile electrons fulfills the so-called H\"uckel rule: $N = 4n_0 +
2$, $n_0 = 1,2, \ldots, 2m_0$. The case $n_0 = m_0$ corresponds to $N = M$ and leads to the half-filled band in the 1D-metal limit. The molecules in the chain (and the
corresponding MOs) will be numbered by indices $m, n, \ldots$ belonging to an $M$-element set
\begin{equation} \label{eq:A(m0)}
{\mathcal A}(m_0) = \{0, \pm 1, \ldots , \pm 2m_0, 2m_0+1\} \, .
\end{equation}
A translation of $n$ nearest-neighbor distance units ($nR^0$, $n = 1, \ldots, M$) along the chain transforms the $m$th molecule into the $(m+n)$th one, where the modulo($M$)
addition of indices belonging to set ${\mathcal A}(m_0)$ is applied. It is implicitly assumed that the environment of any molecule in the X$_M$ system is the same as that
of the $0$th molecule, see Fig.~\ref{fig:cyclic}. Thus, our system is invariant with respect to all the translations (they form a cyclic group of order $M$, hereafter
denoted as $\mathcal{T}_M$). The above construction leads to the X$_M$ system which is {\em cyclic\/}, but {\em locally linear\/}.

  The electronic-structure model of the X$_M$ system, with a single molecular orbital per subsystem, may be easily cast into the form of the PPP model of polyacetylene in which the $2p_z$ atomic orbitals  of the carbon atoms are replaced by the molecular orbitals of the subsystems. The original orbitals of the molecules in the X$_M$ system are nonorthogonal, and the overlap between the orbitals of the adjacent molecules is vital for making the electron transfer along the chain possible. However, the handling of nonorthogonal orbitals in a semiempirical model is inconvenient, and it will be assumed that the set of the original molecular orbitals is subject to the symmetrical orthonormalization procedure of L\"{o}wdin,~\cite{lowdin:50} yielding a set of {\em orthogonalized\/} molecular orbitals which fulfill the {\em orthonormality conditions\/}: 
\begin{equation}
\label{eq:OMOs}
\langle \chi_m |\chi_n \rangle = \delta_{mn} \,.
\end{equation}
These orbitals are well localized and similar to the original molecular orbitals, including the symmetry properties: the translation of $nR^0$ transforms $\chi_m$ into $\chi_{m + n} \,$.  It was Fisher-Hjalmars~\cite{hjalmars:65a,hjalmars:65b} who showed that certain simplifying assumptions of the PPP model for the $\pi$-electron molecules, involving the so-called ZDO approximation introduced by Parr,~\cite{parr:52} can be substantiated when the one- and two-electron integrals, which become empirical parameters in the PPP model, are treated as corresponding to the orthogonalized orbitals. The crucial simplification achieved in the PPP model is due to the complete neglect of all the three- and four-center two-electron integrals, as well as the two-center two-electron integrals of the hybrid and exchange types. Specifically, we assume that for our orthogonalized orbitals of Eq.~(\ref{eq:OMOs}):
\begin{multline}\label{eq:OMO2el}
\langle \chi_m (1) \chi_n(2) | e^2 r_{12}^{-1} \chi_{m'} (1) \chi_{n'} (2) \rangle 
= \\\delta_{mm'} \, \delta_{nn'} \, 
\langle \chi_m (1) \chi_n (2) | e^2 r_{12}^{-1} \chi_m (1) \chi_n (2) \rangle \,,
\end{multline}
where $e$ stands for the elementary charge. In other words, in the PPP model the only surviving two-electron integrals are the two-center integrals of the Coulomb type, as well as the one-center integrals.

The key to the explanation why Eq.~(\ref{eq:OMO2el}) should work lies in the properties of the two-center densities $ \chi_m (1) \chi_n(1)$, $m \neq n$, which contribute to the two-electron integrals neglected in the PPP model. These densities integrate to the corresponding overlap integrals, which vanish for the orthogonalized orbitals that fulfill conditions (\ref{eq:OMOs}). The performance of the approximation (\ref{eq:OMO2el}) may be illustrated by the results obtained for benzene by McWeeny.~\cite{mcweeny:55} He calculated the two-electron integrals corresponding to the orthogonalized basis of atomic orbitals, and found that the absolute values of integrals neglected in the PPP model do not exceed $0.12$~eV, while the values corresponding to the original (nonorthogonal) basis of atomic orbitals are as big as $3.31$~eV (these results are quoted in Parr's book,~\cite{parr:63} see Table 9 on p. 67). Similar behavior may be expected for the orthogonalized basis of molecular orbitals corresponding to a general X$_M$ system considered in the present paper.

The implicit use of the orthogonalized basis in the PPP model, which validates approximation (\ref{eq:OMO2el}), supersedes the old-fashioned ZDO approximation~\cite{parr:52} employed in the original formulation~\cite{pariser:53a,pariser:53b,pople:53} of the model. Let us stress that the approximation given in Eq.~(\ref{eq:OMO2el}) is valid {\em only} when orbitals $\chi_m$ fulfill Eq.~(\ref{eq:OMOs}). The use of a nonorthogonal basis, and a selective incorporation of the corresponding two-electron integrals (as proposed, e.g., in Ref.~\onlinecite{campbell:90}), is likely to spoil the internal consistency of the PPP model.

\subsection{Pariser-Parr-Pople Hamiltonian}
\label{ssec:PPPH}

The second-quantized version of the Pariser-Parr-Pople (PPP) many-electron Hamiltonian was introduced by Kouteck\'{y},~\cite{koutecky:67} see also a paper by Del Re.~\cite{delre:90} In the case of our X$_M$ system, we consider the Fock space spanned by all possible $N$-electron Slater determinants ($N = 0, 1, \ldots , 2M$) built of spin-orbitals from the $2M$-element orthonormal spin-orbital basis set
${\mathcal B} = \{ \chi_{m} \alpha, \chi_{m} \beta \} \,$, where $\alpha$ and $\beta$ are the one-electron spin functions. Any linear operator acting in our Fock space may be built from products of annihilation and creation operators: $\hat{a}_{m \alpha}$ is the annihilation operator associated with the spin-orbital $\chi_m \alpha \,$, and
$\hat{a}_{m \alpha}^{\dagger}$ is the corresponding creation operator. The Fock-space Hamiltonian for our chain, built according to the prescriptions of the PPP model, reads as 
\begin{align}\label{eq:PPPH(1)} \hat{H} & =  \alpha^0 \hat{N} + \beta^0 \sum_{m \in {\mathcal A}(m_0)} (\hat{t}_{m,m+1} + \hat{t}_{m+1,m})\nonumber \\
&  \quad{} + \gamma^0 \sum_{m\in {\mathcal A}(m_0)} 
\hat{n}_{m \alpha} \hat{n}_{m \beta} \nonumber 
\\  & \quad{} + \frac{1}{2} \sum_{m \neq n \in {\mathcal A}(m_0)} \gamma (|n-m| R^0) (z^0 - \hat{n}_m)
(z^0 - \hat{n}_n) \,. \end{align}
The meaning of the operators appearing in Eq.~(\ref{eq:PPPH(1)}) is as follows: for the spin-$\alpha$ electrons $\hat{n}_{m \alpha} =
\hat{a}_{m \alpha}^{\dagger} \hat{a}_{m \alpha}$ is the occupation-number operator corresponding to site $m \,$, and $\hat{t}_{mn \alpha} = \hat{a}_{n \alpha}^{\dagger}
\hat{a}_{m\alpha}$ is the electron-transfer operator corresponding to transfer from site $m$ to $n\,$;  we define also $\hat{n}_m = \hat{n}_{m \alpha} + \hat{n}_{m \beta}
\,$, $\hat{t}_{mn} = \hat{t}_{mn \alpha} + \hat{t}_{mn \beta} \,$, and the electron-number operator reads as $\hat{N} = \sum_m \hat{n}_m \,$.  Quantities $\alpha^0$,
$\beta^0$, and $\gamma^0$ are empirical parameters of the PPP model: $\alpha^0$ represents the binding energy of an electron described by any of the orbitals $\chi_m \,$, $\beta^0$ corresponds to the electron transfer between a pair of the neighboring orbitals, $\chi_m$ and $\chi_{m+1} \,$, and 
\begin{equation}\label{eq:OMO2el1c} 
\gamma^0 = \langle \chi_m (1) \chi_m (2) | 
e^2 r_{12}^{-1} \chi_m (1) \chi_m (2) \rangle 
\end{equation} 
is the one-center two-electron integral. In the case when the outer molecular orbital
is empty, we shall refer
to the molecule as to the (molecular) core; the core plus $z^0$ electrons is assumed to make an electrically neutral system. Parameter $z^0$ should assume a discrete
value: $0,1,$ or $2$, but it will be advantageous to treat it as a continuous quantity, $0 \le z^0 \le 2$, see further discussion. The mobile electrons are assumed to move
in a static potential of molecular cores, and $z^0$ may be treated as representing an {\em effective charge\/} of the molecular core (in the units of the elementary charge
$e$). The last term of Hamiltonian (\ref{eq:PPPH(1)})  represents a sum of the long-range electrostatic interactions of the core-core, electron-core, and electron-electron
types. The dependence of these interactions on the distance $R$ is described in the PPP model by a {\em common\/} function $\gamma(R)$. Here we depart from the PPP model of Ref.~\onlinecite{stolarczyk:88}, where function $\gamma(R)$ describing the electron-electron interactions was different from that used for the core-core and core-electron interactions. 

\begin{figure}
\includegraphics{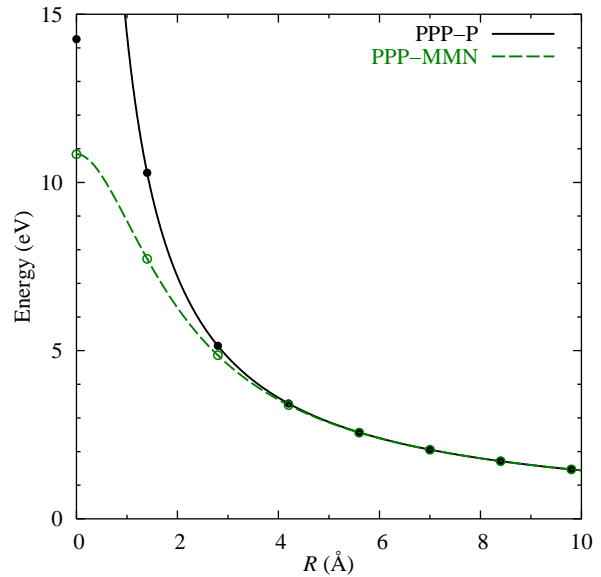}%
\caption{(Color online) \label{fig:gamma(R)}
Functions $\gamma(R)$ corresponding to PPP-P and PPP-MMN models. Values corresponding to $R = n R^0$, $n = 0, 1,2, \ldots$, for $R^0 =1.4$~\AA, are indicated by circles.}
\end{figure}

In Eq. (\ref{eq:PPPH(1)}), function $\gamma(R)$ represents the two-center two-electron integral of the Coulomb type:
\begin{equation}\label{eq:OMO2el2c}
\gamma(R) = \langle \chi_m (1) \chi_n (2) | e^2 r_{12}^{-1} \chi_m (1) \chi_n (2) \rangle \,,
\end{equation} 
in which the centers of molecular orbitals $\chi_m$ and $\chi_n$ are displaced by distance $R = |n-m| R^0 \,$; by definition $ \gamma(0) = \gamma^0\,$. In the PPP model, function $\gamma(R)$ is considered as a certain {\em interpolating function\/} between the one-center value (\ref{eq:OMO2el1c}) and the long-range limiting form, $\gamma(R) \rightarrow {\it e}^2/R$ for $R \rightarrow \infty$. The original proposal by Pople~\cite{pople:53} was to use the bare Coulomb potential for the two-center interactions:
\begin{equation}\label{eq:gammaP(R)}
\gamma_{\rm P} (R) = \left\{ 
\begin{array}{lcr}
  \gamma_{\rm P}^0  & {\mbox {\rm for}} & R = 0 \,, \\ 
  e^2/R              & {\mbox {\rm for}} & R \neq 0 \,. 
\end{array}
\right.
\end{equation}
Although the Pople function $\gamma_{\rm P} (R)$ has never won popularity in the molecular applications of the PPP model, we find it quite a useful reference function in application to the 1D metals, with the one-center parameter $\gamma_{\rm P}^0$ defined as
\begin{equation}\label{eq:gamma0P}
 \gamma_{\rm P}^0 =  (2 \ln 2) e^2/R^0 \,,
\end{equation}
see Sec.~\ref{sec:HF}. 
A general interpolating function in the PPP model may be written as
\begin{equation}\label{eq:gamma(R)}
 \gamma(R) = \gamma_{\rm P} (R) + \lambda(R) \,,    
\end{equation}
 where
\begin{equation}\label{eq:lambda(R)}
\lambda(R) = \left\{ 
\begin{array}{lcr}
  \gamma^0 - (2 \ln 2) e^2/R^0   & {\mbox {\rm for}} & R = 0       \,, \\ 
  \gamma(R) - e^2/R                    & {\mbox {\rm for}} & R \neq 0 \,, 
\end{array}
\right.
\end{equation}
and function $\lambda(R)$ has to decay faster than $1/R$, i.e., $R \, \lambda(R) \rightarrow 0$ for $R \rightarrow \infty$. Obviously, the Pople function (\ref{eq:gammaP(R)}) describes the essential Coulombic part of any interpolating function $\gamma(R)$. When the interacting charge distributions have spherical symmetry, the corresponding function $\lambda(R)$ describes the so-called charge-penetration effects, and should decay exponentially for $R \rightarrow \infty$. Such behavior is desirable in applications to model extended systems, but is {\em not fulfilled\/} by the most popular interpolating functions proposed by Mataga and Nishimoto,~\cite{mataga:57} and by Ohno~\cite{ohno:64}. In Ref.~\onlinecite{stolarczyk:88} the so-called modified Mataga-Nishimoto (MMN) interpolating function was introduced,
\begin{equation}\label{eq:gammaMMN(R)}
 \gamma_{\rm MMN}(R) = e^2 \left[ R + e^2 (\gamma^0)^{-1} \exp({\!} - e^{-2} \gamma^0 R) \right]^{-1} ,    
\end{equation}
which corresponds to an exponentially decaying function $\lambda(R)$. Functions $\gamma_{\rm P}(R)$ and $\gamma_{\rm MMN}(R)$, hereafter also referred to as potentials, are depicted in Fig.~\ref{fig:gamma(R)}. These potentials correspond to the parametrization of the PPP model of Table~\ref{tab:param}, which will be used throughout the paper (it is the same parametrization as
employed for the model of polyacetylene in Ref.~\onlinecite{stolarczyk:88}). 

\begin{table}
\caption{
\label{tab:param}
Parameters of Hamiltonian (\ref{eq:PPPH(1)}) employed in calculations of the present paper. Energies are in eV, and $R^0$ is in \AA; the conversion factors to atomic units are the same as in Refs.~\onlinecite{podeszwa:02a,podeszwa:02c,podeszwa:03}: 1 hartree $=27.2116$~eV, 1 bohr $=0.529\mbox{ }177$~\AA.
}
\begin{ruledtabular}
\begin{tabular}{lcdddc}
                       & \multicolumn{1}{c}{$\alpha^0$} &  \multicolumn{1}{c}{$\beta^0$} &  \multicolumn{1}{c}{$\gamma^0$} & \multicolumn{1}{c}{$R^0$}  \\
\hline
PPP-P            &      $0$         &  -2.5       &  14.259\footnote{Calculated from Eq.~(\ref{eq:gamma0P})}   & 1.4 \\
PPP-MMN      &      $0$         &  -2.5       &  10.840        & 1.4 \\
Hubbard-$0$ &      $0$         &  -2.5       &    5.000        & \multicolumn{1}{c}{---} 
\end{tabular}
\end{ruledtabular}
\end{table}

\subsection{Extended-system limit \label{ssec:Extsyslimit}}
     The idea behind building a cyclic, locally linear
X$_M$ system, and to study its properties in the extended-system limit ($M \rightarrow \infty$), is based on the assumed high symmetry of the latter model, and on the
expected fast convergence of its properties in this limit. The studied properties must be {\em intensive\/} quantities, having finite limiting values: the total energy per
molecule, excitation energies, and electronic-band energies provide important examples. In the Hubbard model (in general, in any model with a finite range of interparticle
interactions) no problem emerges, since all eigenenergies
behave as {\em extensive\/} quantities. In the case of models with long-range interactions, like
the PPP model described by Hamiltonian~(\ref{eq:PPPH(1)}), only the eigenenergies corresponding to the system with no macroscopic net charge behave as extensive
quantities.

  When studying the electronic states in the extended-system limit, one has to select a certain $N$-electron closed-shell ground state ($N = 4n_0 + 2$) as the
reference state. The choice of a particular value of $N$ in our X$_M$ model system may reflect the doping level of a real system, corresponding to the presence of some
electron-donor or electron-acceptor molecules in the vicinity of the X$_M$ system.  Already at the Hartree-Fock level, in order to arrive at the convergent lattice sums in the
expressions for the Hartree-Fock energy (per molecule) and the orbital energies, one has to assume that the unit cell (molecule X in this case) is electrically neutral. For
Hamiltonian~(\ref{eq:PPPH(1)}), this neutrality condition requires that
\begin{equation}\label{eq:z0}
z^0 = N/M \,,
\end{equation}
which in general leads to a fractional core effective charge $z^0$. By applying the condition (\ref{eq:z0}) one may continuously vary the doping level without destroying the
translational symmetry of the system (no explicit inclusion of counterions is necessary). The extended-system limit should then correspond to the joint conditions:
\begin{equation}\label{eq:extsyslim}
M \rightarrow \infty, \qquad  N/M = z^0  = \mbox{\rm const} \,.
\end{equation}
Let us note that parameter $z^0$ corresponds now also to the mean occupation number (the filling level) of the outer MO in molecule X. In the limit~(\ref{eq:extsyslim}),
quantity \mbox{$0 \le z^0/2 \le 1$} denotes also the filling level of the electron-energy band of our 1D metal.

\subsection{Essential parameters of PPP Hamiltonian \label{ssec:EssparPPPH}}
  
  From a formal point of view, the PPP Hamiltonian~(\ref{eq:PPPH(1)}) depends on five parameters,
\begin{equation}\label{eq:PPPH(1a)}
\hat{H} \equiv \hat{H}(\alpha^0,\beta^0,\gamma^0,z^0,R^0) 
= \hat{H}(0,\beta^0,\gamma^0,z^0,R^0) + \alpha^0 \hat{N} \,,
\end{equation}
as well as on the choice of the potential $\gamma(R)$. Actually, $\beta^0$ should be a certain function of $R^0$, but it is more convenient to treat $\beta^0$ as an independent parameter. From the considerations of the previous subsection, it follows that one has to choose the number $N$ of electrons in the ground state of the system. Then, the condition of electrical neutrality fixes the value of parameter $z^0$, see Eq.~(\ref{eq:z0}). Thus, $z^0$ is no longer a variable in Eq.~(\ref{eq:PPPH(1a)}), which reduces the number of independent parameters to four. 
Since $\hat{N}$ commutes with $\hat{H}$, eigenfunctions of Hamiltonian~(\ref{eq:PPPH(1a)}) are independent of parameter $\alpha^0$, while the corresponding eigenenergies still depend on all the four parameters. 
We shall refer to parameters $\beta^0$, $\gamma^0$, and $R^0$ as to the {\em essential\/} parameters of the PPP Hamiltonian~(\ref{eq:PPPH(1)}).  
  
  When the Pople interpolating function (\ref{eq:gammaP(R)}, \ref{eq:gamma0P}) is used in Hamiltonian~(\ref{eq:PPPH(1)}), parameter $\gamma_{\rm P}^0$ depends on $R^0$. One finds in this case that the scaling of the translation parameter, $R^0 \rightarrow \mu R^0$, brings about the following scaling of $N'$-electron eigenenergies corresponding to $\alpha^0 = 0 \,$:
\begin{equation}\label{eq:scaling}
E_j^{N'} (0,\beta^0, \mu^{-1} \gamma_{\rm P}^0, \mu R^0) = \mu^{-1} E_j^{N'}(0, \mu \beta^0, \gamma_{\rm P}^0, R^0) \,;
\end{equation}
here $N'$ denotes the number of electrons which may be different from the number $N$ corresponding to the ground state. Thus, the calculations performed for some fixed $R^0$ may be easily extended to other values of this parameter. It is seen that the eigenfunctions of Hamiltonian~(\ref{eq:PPPH(1)}) with the Pople potential $\gamma_{\rm P}(R)$ depend effectively on a {\em single\/} parameter, $\beta^0$. This feature makes the corresponding PPP model, hereafter abbreviated PPP-P, a convenient {\em reference model\/} in theoretical studies of 1D metals. It seems that for the 1D metals the PPP-P model plays an analogous role to that of the electron-gas model for the 3D metals.

\subsection{Alternancy symmetry and pairing of energy levels \label{ssec:altsym}}

     Let $\Psi_0^N$ represent the closed-shell ground state of the X$_M$ system for the assumed doping level characterized by $z^0$ of Eq. (\ref{eq:z0}). The excited and
ionized states $\Psi_j^{N'}$ correspond to the number of electrons $N' = N, N \pm 1, N \pm 2, \ldots$. By changing parameter $z^0$ to $2 - z^0$, one arrives at a similar
family of $(2M-N')$-electron states. 
It can be shown that due to the alternancy symmetry (see Kouteck\'{y} {\it et
al.}~\cite{koutecky:85} and references therein), for the ground-state eigenenergies per molecule, ${\mathcal E}_0^N = M^{-1} E_0^N$, and for the excitation energies 
$\Delta E_{0,j}^{N,N'} = E_j^{N'} -E_0^N$, one obtains
\begin{equation}\label{eq:E(2M-N)=EN(3)}
{\mathcal E}_0^{2M-N} = {\mathcal E}_0^N + (1 - z^0)(2 \alpha^0 + \gamma^0) \,,
\end{equation}
\begin{equation}
\label{eq:E(2M-N)=EN(4)}
 \Delta E_{0,c(j)}^{2M-N,2M-N'} = \Delta E_{0,j}^{N,N'} + (N - N')(2 \alpha^0 + \gamma^0) \,,
\end{equation}
which hold for finite $M$, as well as in the extended-system limit. 
In the above equation, $j \rightarrow {c(j)}$ denotes a certain mapping (pairing) of the indices labeling the $N$- and $(2M-N)$-electron states; for the respective
ground states, one has $c(0) = 0$.
Therefore, one may restrict the study of the X$_M$ system to the filling levels $0 < z^0/2 \le 1/2$.
Eqs.~(\ref{eq:E(2M-N)=EN(3)}) and (\ref{eq:E(2M-N)=EN(4)}) are 
fulfilled also when the exact eigenenergies are replaced by the results of certain approximate calculations,
e.g., by the Hartree-Fock  or the coupled-cluster method. 
The ground-state energies per electron may be calculated as ${\mathcal E}_0^N/z^0$. \\

   Concluding this Section, let us look again at the structure of the PPP Hamiltonian~(\ref{eq:PPPH(1)}). By the neglect of certain terms, the form of this Hamiltonian may
be reduced to that corresponding to a more simplified model. When only the first two terms are retained, one arrives at the second-quantized version of the H\"uckel
Hamiltonian,~\cite{huckel:31} with no explicit electron-electron interactions. By including the third term, the Hubbard Hamiltonian~\cite{hubbard:63} is obtained, where
the electron-electron interactions are confined to individual molecular centers (because of the assumption of the single molecular orbital per center, only the electrons with the
opposite spins may interact). The Hubbard model tries to account for a dynamical screening of the electron-electron interactions (which is an electronic-correlation
effect) by actually purging out {\em all\/} the long-range terms from the Hamiltonian (including the electron-core and the core-core terms, which have to be sacrificed for
the sake of the electrostatic balance). There is also a family of the so-called extended Hubbard models, in which the long-range electrostatic terms are neglected after
the $n$ neighbors ($n= 1, 2, \ldots$); we shall refer to these models as to the Hubbard-$n$ models, the original Hubbard model being Hubbard-$0$. In Table~\ref{tab:param}
we give also the parameters corresponding to the Hubbard model (for annulenes). A low value for the one-center two-electron integral in this model ($\gamma^0_{\rm H} =
5.0$~eV) was suggested by Paldus and Boyle,~\cite{paldus:82} who argued that $\gamma^0_{\rm H}$ should be approximately equal to the difference $\gamma^0 - \gamma(R^0)$
corresponding to the PPP model.~\cite{paldus:74}

\section{\label{sec:HF}%
Hartree-Fock results for 1D-metal model}

The basis set of orthogonalized molecular orbitals $\{ \chi_m \,, m \in {\mathcal A}(m_0) \}$ is in one-to-one correspondence with the orthonormal basis set of symmetry orbitals (the Bloch orbitals)
corresponding to the translation group ${\mathcal T}_M$:
\begin{equation}\label{eq:COs}
\psi_k = M^{-1/2} \sum_{m \in {\mathcal A}(m_0)}
\chi_m \, \exp \left( \frac{2 \pi i}{M} k  m \right) \,,
\end{equation}
where $k \in {\mathcal A}(m_0)$ enumerates the symmetry labels of
$M$ one-dimensional  representations of the ${\mathcal T}_M$  group, which are
complex for $k \neq 0, 2m_0 + 1$. For {\em arbitrary\/} state of our X$_M$ system, described within the restricted Hartree-Fock (RHF) theory (in which no symmetry breaking
is allowed), the HF orbitals may be taken in the form of the symmetry orbitals (\ref{eq:COs}). These orbitals will be hereafter referred to as the crystal orbitals (COs).  
The two-electron integrals corresponding to the CO basis, calculated by using Eqs.~(\ref{eq:OMO2el}) and (\ref{eq:OMO2el2c}), may be written as
\begin{equation}\label{eq:CO2el}
\langle \psi_{k_1+q} (1) \psi_{k_2-q} (2) | e^2 r_{12}^{-1} \psi_{k_1} (1) \psi_{k_2} (2) \rangle = v(q) \,,
\end{equation}
where
\begin{equation}\label{eq:v(q)}
v(q) =  v(-q) = M^{-1} \sum_{m \in {\mathcal A}(m_0)} 
\gamma(m R_0) \, \cos \left( \frac{2 \pi}{M} \, q  m \right) \,.
\end{equation}
In the above expressions the symmetry labels $k_1, k_2, q \in {\mathcal A}(m_0)$ are subject to the modulo($M$) addition rule. It is seen that in the PPP model of the 1D
metal the two-electron CO integrals are enumerated by a {\em single} index $q \in {\mathcal A}(m_0)$.

   We are going to study the closed-shell ground state of the X$_M$ system described by the PPP Hamiltonian~(\ref{eq:PPPH(1)}). We assume that the number of electrons
conforms to the H\"{u}ckel rule: $N = 4n_0 +2$, $n_0 = 1,2, \ldots \,$, and that condition~(\ref{eq:z0}) holds. The $N$-electron ground state of the system, $\Psi_0^{N}$,
corresponds to the spin singlet, the fully symmetric representation ($k = 0$) of the translation group, and, for $N = M$ it belongs to the ``minus'' category corresponding
to the alternancy symmetry.~\cite{pariser:56}
Throughout this paper we shall assume that the overlap integral between the original molecular orbitals of the adjacent molecules
is $ > 0$, and thus parameter $\beta^0 < 0$. In such a case the RHF determinantal wave function $\Phi_{\rm HF}$ ($\equiv \Phi_0^{N}$) describing the ground state is
characterized by a double occupation of the COs with the symmetry label $k$ belonging to the set
\begin{equation}\label{eq:O(n0)}
{\mathcal O}(n_0) = \{0, \pm 1, \ldots , \pm n_0 \} \,.
\end{equation}
The above set contains $2n_0 + 1 = N/2$ elements and will be referred to as the occupied CO index set. The $(M-N/2)$-element subset of ${\mathcal A}(m_0)$ which is
complementary to ${\mathcal O}(n_0)$ reads as
\begin{equation}\label{eq:U(m0,n0)}
{\mathcal U}(m_0,n_0) =  \{ \pm (n_0 +1), \pm (n_0 +2), 
\ldots  , \pm 2m_0, 2m_0 + 1 \} \,,
\end{equation}
and will be referred to as the unoccupied CO index set. The CO energies of the X$_M$ system are calculated as
\begin{align}\label{eq:epsHF(k)}
\varepsilon_{\rm HF}(k) & =  \langle \psi_k |\hat{f} \psi_k \rangle \nonumber \\
           & =  \alpha^0 + 2 \beta^0 \cos \left( \frac{2 \pi}{M} \, k\right) \nonumber\\
&\quad{} + \gamma^0 \, z^0 - \sum_{q \in {\mathcal O}(n_0)}  v(k - q) \,,
\end{align}
where $\hat{f}$ is the Fock operator corresponding to $\Phi_{\rm HF}$ and  Hamiltonian~(\ref{eq:PPPH(1)}). The CO energies $\varepsilon_{\rm HF}(k) \equiv \varepsilon_{\rm
HF}^{N}(k)$ depend on the parameters $\alpha^0,\beta^0,\gamma^0,z^0,$ and $R^0$.
Due to the time-reversal symmetry, the complex-conjugate COs
$\psi_k$ and $\psi_{-k}$ are degenerate: $\varepsilon_{\rm HF}^{N}(-k) = \varepsilon_{\rm HF}^{N}(k)$. In the one-electron approximation, the alternancy symmetry 
manifests itself through the Coulson-Rushbrooke theorem,~\cite{coulson:40} which introduces another pairing among the COs: $\psi_k$ and $\psi_{k -
q_0}$, where $q_0 = 2m_0 +1$, make a conjugate pair. It can be shown in general, by invoking Koopmans' theorem~\cite{koopmans:34} 
and using formula
(\ref{eq:E(2M-N)=EN(4)}), 
that the CO energies corresponding to the $N$- and $(2M - N)$-electron ground-state RHF functions are related to each other:
\begin{equation}\label{eq:altsymepsHF}
\varepsilon_{\rm HF}^{2M - N}(k)  - (\alpha^0 + \gamma^0/2) 
= - \big[ \varepsilon_{\rm HF}^{N}(k - q_0)  - (\alpha^0 + \gamma^0/2) \big] \,.
\end{equation}
If one of the paired one-electron states corresponding to the above formula  is occupied, then the other one is unoccupied. In particular, for $N = M$, the alternancy symmetry
causes the pairing of the occupied and unoccupied COs, and their orbital energies satisfy Eq. (\ref{eq:altsymepsHF}).

   The HF energy per molecule, ${\mathcal E}_{\rm HF} = {\mathcal E}_{\rm HF}^N$, in the the X$_M$ system reads as
\begin{align}\label{eq:EHF}
{\mathcal E}_{\rm HF} & =  M^{-1} \langle \Phi_{\rm HF} |\hat{H} \Phi_{\rm HF} \rangle \nonumber \\
                        & =  \alpha^0 \, z^0 + \beta^0 \frac{4 \sin( z^0 \pi /2)} {M \sin(\pi /M)} \nonumber \\
		&\quad {}+ \gamma^0 \, \frac{(z^0)^2}{2} 
- M^{-1} \sum_{k,q \in {\mathcal O}(n_0)} v(k - q) \,.
\end{align}
The terminal terms at the rhs of Eqs.~(\ref{eq:epsHF(k)}) and (\ref{eq:EHF}) correspond to the nonlocal Hartree-Fock exchange. In the Hubbard model, these terms reduce to $-\gamma^0
\, z^0/2 $ and $-\gamma^0 \, (z^0)^2/4 $, respectively (these are the terms correcting for the one-center self-interaction of electrons).

For $N \neq 0, 2M$, the gap between the degenerate pair of the highest occupied COs and the degenerate pair of the lowest unoccupied COs amounts to 
\begin{align}\label{eq:epsHFgap}
\Delta \varepsilon_{\rm HF} &= \varepsilon_{\rm HF}(n_0 + 1) - \varepsilon_{\rm HF}(n_0)  \nonumber\\
&= - 4 \beta^0 \, \sin ( z^0 \pi /2) \sin (\pi /M) + v(0) -  v(N/2) \,.
\end{align}
Interestingly, for finite $M$ the above gap is always $> 0$, and does not vanish even for $\beta^0 = 0$, due to the exchange term in $\varepsilon_{\rm HF}(k)$.

   When studying the extended-system limit of the X$_M$ system, it will be advantageous to replace the integer symmetry labels $k \in  {\mathcal  A}(m_0)$ by 
\begin{equation}\label{eq:underlinek}
\underline{k} =  \frac{2 \pi}{M} \, k \,,  
\end{equation}
where the new labels form an evenly spaced $M$-element subset in the semiclosed interval $(- \pi, \pi]$. The new symmetry label $\underline{k}$ is proportional to the
quasimomentum $\underline{k} \hbar / R^0$, where $\hbar$ is the Planck constant, and will be for brevity referred to as the quasimomentum. The addition rule for the
quasimomenta $\underline{k} \in (- \pi, \pi]$ is borrowed from the modulo($M$) addition rule in set ${\mathcal A}(m_0)$; this corresponds to the so-called Umklapp process
for the quasimomentum. As a consequence, arbitrary function of quasimomentum, $f(\underline{k})$, which is defined for $\underline{k} \in (- \pi, \pi]$, may be extended to
the whole range of real arguments by assuming that it is periodic with period $2\pi$; this interpretation of the Umklapp process will be used throughout the paper. The
quasimomenta corresponding to the occupied COs, with labels $k \in {\mathcal O}(n_0)$, form an $N/2$-element subset of the closed interval $[-k_{\rm F}, k_{\rm F}]$, where
we introduce the so-called Fermi quasimomentum:
\begin{equation}\label{eq:kF}
k_{\rm F} =  \frac{\pi N}{2 M} =  z^0 \frac{\pi}{2} \,.  
\end{equation}
The Fermi quasimomentum $k_{\rm F}$ is related to the effective core charge $z^0$ because of condition (\ref{eq:z0}). Interval $(- \pi, \pi]$ represents the first
Brillouin zone for our 1D metal, and $[-k_{\rm F}, k_{\rm F}]$ is the Fermi interval (the 1D analog of the Fermi sphere). The quasimomenta corresponding to the unoccupied
COs, with labels $k \in {\mathcal U}(m^0,n_0)$, form an $(M-N/2)$-element subset belonging to the union of intervals $(-\pi, -k_{\rm F})$ and $(k_{\rm F}, \pi]$.

    In the limit of $M \rightarrow \infty$, the quasimomenta $\underline{k}$ related to sets ${\mathcal A}(m_0)$, ${\mathcal O}(n_0)$, and ${\mathcal U}(m_0, n_0)$ form
dense countable subsets enclosed in the respective intervals. In this limit, a summation over symmetry labels forming a subset in ${\mathcal A}(m_0)$ may be replaced by an
integration over the corresponding subinterval in $(-\pi, \pi]$:
\begin{equation}\label{eq:sum-to-int}
M^{-1} \sum_{k=k_1}^{k_2}  \rightarrow
 (2\pi)^{-1} \int_{\underline{k}_1}^{\underline{k}_2} d \underline{k}  \,,
\end{equation}
where $-2m_0 \le k_1 \le k_2 \le 2m_0+1$. For simplicity, in the formulas corresponding to the infinite limit we shall replace the symbol of quasimomentum,
$\underline{k}$, by symbol $k$, wherever no confusion arises. In the extended-system limit (\ref{eq:extsyslim}) we shall use the Fermi quasimomentum (\ref{eq:kF}) rather
than parameter $z^0$ to characterize the reference ground state of the system.

\begin{figure}
\includegraphics{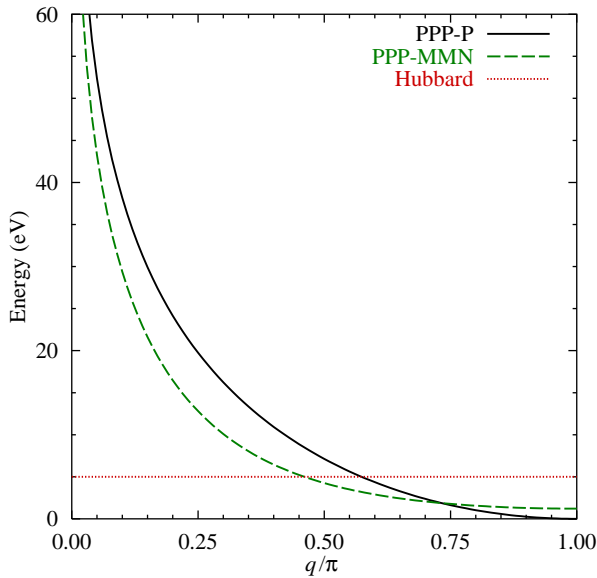}%
\caption{(Color online) \label{fig:V(q)}
Functions $V(q)$ corresponding to PPP-P, PPP-MMN, and Hubbard-0 models.}
\end{figure}

    Let us write the two-electron CO integrals (\ref{eq:CO2el}) as
\begin{equation}\label{eq:CO2el(2)}
v(q) = M^{-1} V(q) \,,
\end{equation}
where $q$ now stands for the quasimomentum, see Eq. (\ref{eq:underlinek}), $q \in (-\pi,\pi]$. In the next step we apply partition (\ref{eq:gamma(R)}) to formula
(\ref{eq:v(q)}), and assume that function $\lambda(R)$ vanish beyond $R \ge L_0 R^0$, where $L^0 >0$ is some finite integer. For $M > 2 L_0$ function $V(q)$ may be then
written as
\begin{equation}\label{eq:V(q)}
V(q) = V_{\rm P}(q) +  \lambda(0) + 2 \sum_{m = 1}^{L_0} \lambda(m R^0) \cos m q  \,,
\end{equation}
where $V_{\rm P}(q)$ corresponds to the Pople function defined in Eq. (\ref{eq:gammaP(R)}),
\begin{equation}\label{eq:VP(q)}
V_{\rm P}(q) = \frac{e^2}{R^0} \left\{ 2 \ln 2 
+ 2 \sum_{m = 1}^{2m_0} \frac{\cos m q}{m} 
+ \frac{\cos[(2m_0 + 1)q]}{2m_0 + 1} \right\} \,.
\end{equation}
When studying the infinite-$M$ limit of function $V(q)$, one finds that the second term at the rhs of Eq.~(\ref{eq:V(q)}) reaches its limiting value for $M > 2 L_0$. On
the other hand, the lattice sum of the cosines in $V_{\rm P}(q)$ cannot be effectively summed up by a finite-summation method, see Ref.~\onlinecite{stolarczyk:88} and
references therein. Fortunately, an analytical limiting formula is available:~\cite{stolarczyk:88}
\begin{equation}\label{eq:VP(q)(2)}
V_{\rm P}(q) = - \frac{2 e^2}{R^0} \ln |\! \sin(q/2) | \,,
\end{equation}
for $q \in (-\pi, \pi]$. Thanks to a special choice of the one-center contribution, Eq. (\ref{eq:gamma0P}), the resulting formula is very compact and corresponds to a
nonnegative function. Let us note that $V_{\rm P}(q)$ vanish for $q = \pi$ and diverges logarithmically for $q \rightarrow 0$. Function $V_{\rm P}(q)$ and function $V_{\rm
MMN}(q)$, corresponding to the modified Mataga-Nishimoto function (\ref{eq:gammaMMN(R)}), are depicted in Fig.~\ref{fig:V(q)}. In addition, we show there also a (constant) function $V_{\rm H}(q) = \gamma^0_H$ corresponding to the Hubbard-$0$ model.

The logarithmic divergence of $V(q)$ at $q = 0$ is an inherent feature the PPP model of the 1D metal. It originates from the asymptotic $R^{-1}$ behavior of function
$\gamma(R)$, and is thus absent in the Hubbard-$n$ model for any finite $n$ (for a comparison of the PPP and the Hubbard-1 models of the infinite polyene, see
Ref.~\onlinecite{stolarczyk:88}). In the PPP model one finds also that the CO Coulomb integral $v(0)$, see Eq.~(\ref{eq:CO2el}), vanishes as $(2 e^2/R^0) \ln(M)/M$ in the
limit of $M \rightarrow \infty$. Thus, so vanishes the HF orbital-energy gap (\ref{eq:epsHFgap}) in this limit. Formula (\ref{eq:V(q)}) is applicable when $\lambda (R)$ is
a quickly decaying function. However, when the charge distribution corresponding to (electrically neutral) molecule X has a dipole or quadrupole moment (or when X$_M$
forms a zig-zag chain), $\lambda (R)$ should include terms decaying like $R^{-n}$, $n > 1$. In such a case the necessary lattice sums may be handled by using a general
method described in Refs.~\onlinecite{stolarczyk:88} and \onlinecite{jeziorska:90}.

  In order to find the formulas for the CO energies (\ref{eq:epsHF(k)}) and the HF energy per molecule (\ref{eq:EHF}) in the extended-system limit (\ref{eq:extsyslim}), we
use some of the results of the Appendix A of Ref.~\onlinecite{stolarczyk:88}. By writing formula (\ref{eq:VP(q)(2)}) as
\begin{equation}\label{eq:VP(q)(3)}
V_{\rm P}(q) = \frac{2 e^2}{R^0} \big[ \ln 2 + f_1(q) \big] \,,
\end{equation}
and applying the continuous variables (\ref{eq:underlinek}) and the summation-to-integration switch (\ref{eq:sum-to-int}), we arrive at the following formulas:
\begin{align}\label{eq:epsHF(k)(2)}
\varepsilon_{\rm HF}(k) & =  \alpha^0 + 2 \beta^0 \cos k 
+ \gamma^0 \, \frac{k_{\rm F}}{\pi}  \nonumber \\
&\quad {}- \frac{e^2}{\pi R^0} \, \big[ g_2(k + k_{\rm F}) - g_2(k - k_{\rm F}) \big] \nonumber \\
& \quad{}-   \frac{2}{\pi} \sum_{m = 1}^{L_0} \frac{\lambda(m R^0)}{m} \, \sin m k_{\rm F} 
\cos m k \,,
\end{align}
for $k \in (-\pi, \pi]$, and
\begin{align}\label{eq:EHF(2)}
{\mathcal E}_{\rm HF}  & =  \alpha^0 \frac{2k_{\rm F}}{\pi} 
+  \beta^0 \frac{4 \sin k_{\rm F}}{\pi}  + \gamma^0 \left( \frac{k_{\rm F}}{\pi} \right)^2   \nonumber \\
&\quad {}+  \frac{e^2}{\pi^2 R^0} \, \big[ f_3(2k_{\rm F}) - f_3(0) \big]   \nonumber \\
& \quad {}-    \frac{1}{\pi^2} \sum_{m = 1}^{L_0} \frac{\lambda(m R^0)}{m^2} \,
 \sin^2 m k_{\rm F} \,.
\end{align}
where functions $f_1$, $g_2$, and $f_3$ are defined in the Appendix A of Ref.~\onlinecite{stolarczyk:88}. These functions may easily be calculated  by using the analytical
approximants introduced in the Appendix B of Ref.~\onlinecite{stolarczyk:88}. In both Eqs.~(\ref{eq:epsHF(k)(2)}) and~(\ref{eq:EHF(2)}), the first three terms at the rhs
correspond to the Hubbard-$0$ model; by adding the fourth term one arrives at the results of the PPP-P model. In the extended-system limit the dependence of various
quantities on the number of electrons $N$ should be replaced by the dependence on the Fermi quasimomentum $k_{\rm F}$ defined in Eq.~(\ref{eq:kF}), thus $\varepsilon_{\rm
HF}(k) \equiv \varepsilon_{\rm HF}^{k_{\rm F}}(k)$ and ${\mathcal E}_{\rm HF} \equiv {\mathcal E}_{\rm HF}^{k_{\rm F}}$. Equation~(\ref{eq:altsymepsHF}), reflecting the
influence of the alternancy symmetry on the CO energies, now reads as
\begin{equation}\label{eq:altsymepsHF(2)}
\varepsilon_{\rm HF}^{\pi - k_{\rm F}}(k)  - (\alpha^0 + \gamma^0/2) 
= - \big[\varepsilon_ {\rm HF}^{k_{\rm F}}(k - \pi)  - (\alpha^0 + \gamma^0/2) \big] \,,
\end{equation}
where $k \in (-\pi, \pi]$.

\begin{figure}
\includegraphics{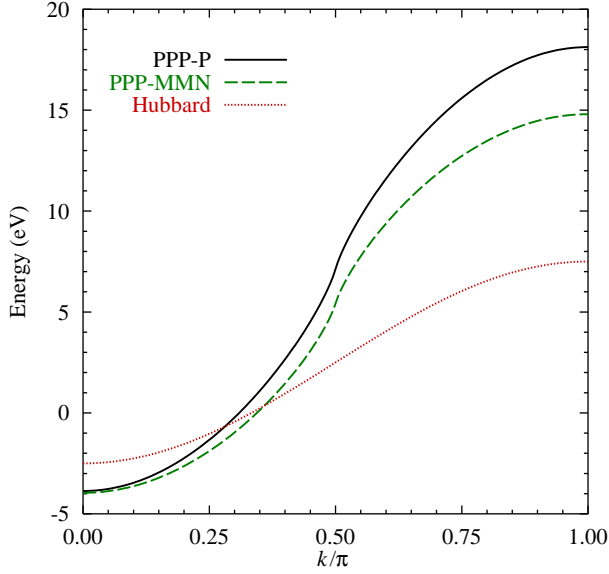}%
\caption{(Color online) \label{fig:epsHF(k)pi/2}
$k_{\rm F} = \pi/2$. Hartree-Fock electronic-energy bands $\varepsilon_{\rm HF}(k)$ for PPP-P, PPP-MMN, and Hubbard-$0$ models.}
\end{figure}

\begin{figure}
\includegraphics{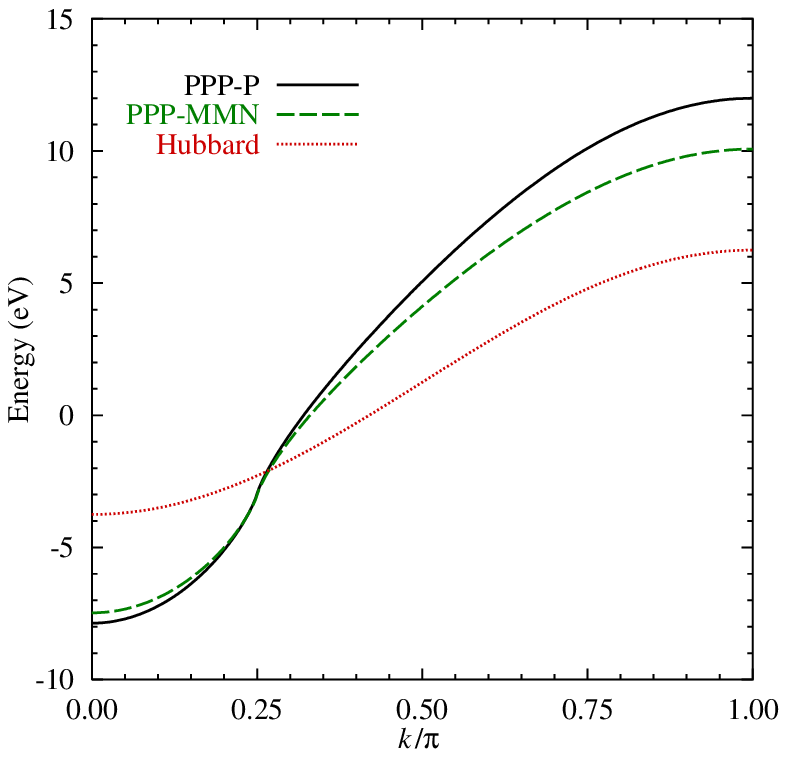}%
\caption{(Color online) \label{fig:epsHF(k)pi/4}
$k_{\rm F} = \pi/4$. Hartree-Fock electronic-energy bands $\varepsilon_{\rm HF}(k)$ for PPP-P, PPP-MMN, and Hubbard-$0$ models.}
\end{figure}

The CO energies given by formula (\ref{eq:epsHF(k)(2)}) represent the electronic-energy bands of our 1D-metal model within the Hartree-Fock approximation.
The first derivative of these CO energies can be calculated in an analytical form; for the PPP-P model one finds that
\begin{multline}\label{eq:derivepsHF(k)}
\frac{d \,\varepsilon_{\rm HF}(k)}{d \, k} = - 2 \beta^0 \sin k 
+ \frac{e^2}{\pi R^0} \big\{
  \ln \! \left| \sin \! \big[ (k + k_{\rm F})/2 \big] \right| \\
- \ln \! \left| \sin \! \big[ (k - k_{\rm F})/2 \big] \right| \big\} \,.
\end{multline}
As pointed out in Ref.~\onlinecite{stolarczyk:88}, the HF-exchange contribution to the CO energies, originating from the long-range Coulombic interactions of electrons, is causing a logarithmic divergence of quantity (\ref{eq:derivepsHF(k)}) at $k = \pm k_{\rm F}$. This in turn makes the density-of-states (DOS) functions vanish at the Fermi level equal to $\varepsilon_{\rm HF}(k_{\rm F})$.~\cite{monkhorst:79,stolarczyk:88} The CO energies corresponding to $k_{\rm F} = \pi/2$ (the half-filled-band case) and $k_{\rm F} = \pi/4$ (the quarter-filled-band case) are shown in Figs.~\ref{fig:epsHF(k)pi/2} and \ref{fig:epsHF(k)pi/4}, respectively. The infinite slope of the energy bands for $k = k_{\rm F}$ is noticeable for the PPP-P and PPP-MMN models, in difference to the results of the Hubbard-$0$ model. The energy bands corresponding to the PPP model are much wider than those corresponding to the Hubbard-$0$ model (in the latter case the total band width amounts simply to $4|\beta^0|$). The differences between the PPP and Hubbard-$0$ energy bands are due solely to the exchange effects that originate from the long-range Coulomb interactions. The vanishing of the DOS function at the Fermi level, and a too-large band width, are well-known pathologies of the HF approximation applied to a metallic system.

\begin{figure}
\includegraphics{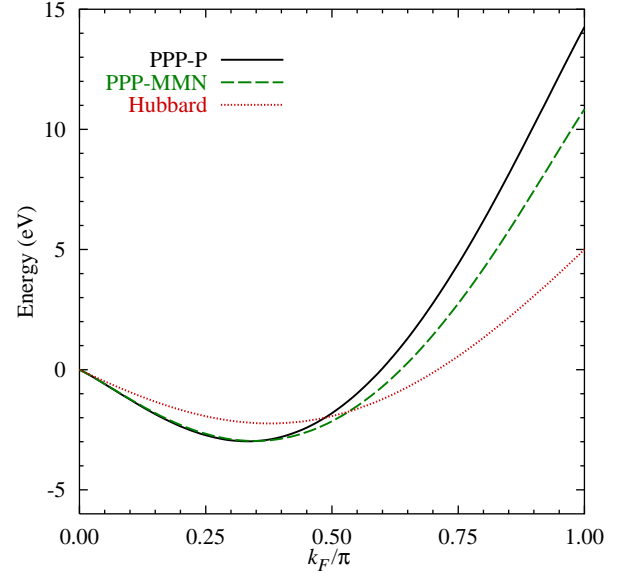}%
\caption{(Color online) \label{fig:EHF(kF)}
Hartree-Fock energy per molecule, ${\mathcal E}_{\rm HF}$, as a function of $k_{\rm F}/\pi$ for PPP-P, PPP-MMN, and Hubbard-$0$ models.}
\end{figure}

   The HF energy per molecule defined in Eq.~(\ref{eq:EHF(2)}) is a function of the Fermi momentum, ${\mathcal E}_{\rm HF} \equiv {\mathcal E}_{\rm HF}^{k_{\rm F}}$; the HF energy per electron may be calculated as ${\mathcal E}_{\rm HF}^{k_{\rm F}} \pi/(2 k_{\rm F})$. The dependence of ${\mathcal E}_{\rm HF}$ on $k_{\rm F }/\pi \in [0,1]$, for the PPP-P, PPP-MMN, and Hubbard-0 models, is depicted in Fig.~\ref{fig:EHF(kF)}. For $\alpha^0 = 0$ (see Table~\ref{tab:param}) and $k_{\rm F} \in [0,\pi/2]$ the value of ${\mathcal E}_{\rm HF} \equiv {\mathcal E}_{\rm HF}^{k_{\rm F}}$ may be viewed upon as corresponding to a delocalization effect (at the HF level): it is equal to the energy lowering with respect to a reference state of the perfectly localized electrons (with no more than one electron per molecule). Near to the half filling, the HF energies corresponding to the Hubbard-$0$ are close to the PPP results, thus providing some justification (at the HF level) for using a smaller parameter $\gamma^0$ in the Hubbard-$0$ model, see Table~\ref{tab:param} and Ref.~\onlinecite{paldus:82}.

   By applying formula (\ref{eq:E(2M-N)=EN(3)}) to the HF energies in the extended-system limit, one finds that
\begin{equation}\label{eq:E(pi-KF)=EkF}
{\mathcal E}_{\rm HF}^{\pi - k_{\rm F}} = {\mathcal E}_{\rm HF}^{k_{\rm F}} 
+ (1 - 2 k_{\rm F} /\pi)(2 \alpha^0 + \gamma^0) \,.
\end{equation}
For specific values of $k_{\rm F}$, compact analytical formulas can be obtained for the HF energies (\ref{eq:EHF(2)}) corresponding to the PPP-P model,  by making use of
Eqs.~(A.5a) and (A.7a) of Ref.~\onlinecite{stolarczyk:88}:
\begin{equation}\label{eq:EHF(3a)}
{\mathcal E}_{\rm HF}^{\pi/2}  =  \alpha^0 +  \beta^0 \frac{4}{\pi} 
+  \frac{e^2}{R^0} \, \left[\frac{\ln 2}{2} - \frac{7}{4 \pi^2} \, \zeta(3) \right] \,,
\end{equation}
\begin{equation}\label{eq:EHF(3b)}
{\mathcal E}_{\rm HF}^{\pi/4} = \alpha^0/2 +  \beta^0 \frac{2 \sqrt{2}}{\pi} 
+  \frac{e^2}{R^0} \, \left[\frac{\ln 2}{8} - \frac{35}{32 \pi^2} \, \zeta(3) \right] \,,
\end{equation}
where $\zeta(3) = 1.202\mbox{ }057$, 
see Ref.~\onlinecite{abramowitz:68}. The value of ${\mathcal E}_{\rm HF}^{3\pi/4}$ may be obtained from Eqs.~(\ref{eq:E(pi-KF)=EkF}) and (\ref{eq:EHF(3b)}). The above results may be extended to any variant of the PPP model by explicitly calculating the last term in Eq.~(\ref{eq:EHF(2)}).

\section{\label{sec:E2tot}%
Second-order correlation corrections to the total energy of 1D-metal model}

The basic many-body techniques, the perturbation theory~\cite{moller:34,brueckner:55,goldstone:57} and the coupled-cluster method,~\cite{coester:58,cizek:66} use as the reference
some determinantal function (most often the Hartree-Fock one). Before such techniques are applied to Hamiltonian (\ref{eq:PPPH(1)}), it is convenient to write it in a different form. By applying a unitary transformation converting the spin-orbital basis ${\mathcal B} = \{ \chi_m \alpha, \chi_m \beta \}$ into ${\mathcal B'} = \{\psi_{k} \alpha, \psi_{k} \beta \}$, and choosing the restricted Hartree-Fock  function $\Phi_{\rm HF}$ as the Fermi vacuum, one may introduce a new representation of linear operators in the Fock space for our X$_M$ system. The PPP Hamiltonian~(\ref{eq:PPPH(1)}) assumes then a general form of the many-electron Hamiltonian:
\begin{equation}\label{eq:PPPH(2)}
\hat{H} = \langle \Phi_{\rm HF} |\hat{H} \Phi_{\rm HF} \rangle + \sum_{p}
\varepsilon_{p} N[\hat{p}^{\dagger} \hat{p}] + \frac{1}{4} \sum_{p,q,r,s}
v^{pq}_{rs} N[\hat{p}^{\dagger} \hat{q}^{\dagger} \hat{s} \hat{r}] \,,
\end{equation}
where $p,r,s$, and $t$ stand for composite spin-orbital
indices: $p = (k, \sigma)$, where $\sigma$ is spin index (equal to
$\alpha$ or $\beta$), and $\hat{p}$ and $\hat{p}^{\dagger}$ are the
annihilation and creation operators, respectively, corresponding to
spin-orbital $\psi_k \sigma$. $N[\cdots]$ stands for the normal
ordering of the enclosed operators with respect to the Fermi vacuum $\Phi_{\rm HF}$. Quantities $ v^{pq}_{rs}$ are the antisymmetrized two-electron integrals
calculated in the spin-orbital basis ${\mathcal B'}$:
\begin{align}\label{eq:vpqrs}
 v^{p_3, p_4}_{p_1, p_2}
&= v^{(k_1 + q) \sigma_3, \, (k_2 - q) \sigma_4}_{k_1 \sigma_1,\; k_2
\sigma_2} \nonumber \\
&= v(q) \, \delta_{\sigma_1,\sigma_3} \,
\delta_{\sigma_2,\sigma_4}
-  v(k_2 -k_1 - q) \, \delta_{\sigma_1,\sigma_4} \,
\delta_{\sigma_2,\sigma_3} \,,
\end{align}
where $v(q)$ are the CO two-electron integrals defined in Eqs.~(\ref{eq:CO2el}) and (\ref{eq:v(q)}). The HF spin-orbital energies $\varepsilon_{p} \equiv \varepsilon_{k, \sigma} = \varepsilon_{\rm HF}(k)$ are equal to the CO energies of Eq. (\ref{eq:epsHF(k)}).

  A modern approach to the many-body perturbation theory places it in the framework of the coupled-cluster (CC) theory.~\cite{bartlett:81,monkhorst:81} In the present paper, we shall pursuit this connection to put our perturbative results into a broader context.  In the CC theory, the eigenfunction of Hamiltonian~(\ref{eq:PPPH(2)}), corresponding to the $N$-electron spin-singlet ground state $\Psi_0 \equiv \Psi_0^N$ with eigenenergy $E_0 \equiv E_0^N $, can be expressed as
\begin{equation}\label{eq:exp(T)}
\Psi_0 = \exp(\hat{T}) \Phi_{\rm HF} \,,
\end{equation}
where $\hat{T}$ is the CC operator, being a sum of the single, double, \ldots{}, up to $N$-tuple excitation operators. The $n$-tuple excitation operator $\hat{T}_n$ depends on some linear parameters, called the $t_n$ amplitudes. The ground-state electronic-correlation energy, $E_{\rm corr} = E_0 - \langle \Phi_{\rm HF} |\hat{H} \Phi_{\rm HF} \rangle$, can be expressed as a simple function of the $t_1$ and $t_2$ amplitudes and the two-electron integrals (\ref{eq:vpqrs}). For the 1D-metal model of Sec.~\ref{sec:PPP}, the COs are determined by the translational symmetry, and the HF function $\Phi_{\rm HF}$ is equivalent to the Brueckner (or maximum-overlap) determinantal function, see Ref.~\onlinecite{stolarczyk:84} and references therein. Therefore, the $t_1$ amplitudes vanish by symmetry, and the electronic-correlation energy per molecule for our X$_M$ system depends only on the $t_2$ amplitudes of operator $\hat{T}_2$:
\begin{equation}\label{eq:T2}
\hat{T}_2=  \frac{1}{4} \sum_{ij}  \sum_{ab} t^{ab}_{ij} 
N[\hat{a}^\dagger \hat{b}^\dagger \hat{\jmath}\hat{\imath}] \,,
\end{equation}
cf. Eq.~(\ref{eq:PPPH(2)}), where indices $i,j$ enumerate the occupied, and $a,b$ -- the unoccupied spin-orbitals. For our 1D-metal model, the $t_2$ amplitudes
corresponding to operator (\ref{eq:T2}) can be expressed as~\cite{podeszwa:02a}
\begin{align}\label{eq:tabij}
 t^{a_3, a_4}_{i_1, i_2} &=  
t^{(k_1 + q) \sigma_3, \, (k_2 - q) \sigma_4}_{k_1 \sigma_1,\; k_2 \sigma_2}  \nonumber\\
&=  
t(k_1, k_2, q) \, \delta_{\sigma_1,\sigma_3} \, \delta_{\sigma_2,\sigma_4}\nonumber \\
&\quad{}-  t(k_1, k_2, k_2-k_1- q) \, \delta_{\sigma_1,\sigma_4} \, \delta_{\sigma_2,\sigma_3} \,,
\end{align}
cf.~Eq.~(\ref{eq:vpqrs}), where (real) quantities $t(k_1, k_2, q)$ may be regarded as the nonorthogonally spin-adapted $t_2$ amplitudes (see, e.g., Ref.~\onlinecite{stolarczyk:84}). The spin-adapted amplitudes $t(k_1, k_2, q) = t(k_2, k_1, -q)$ correspond to the same double excitation. In Eq.~(\ref{eq:tabij}) indices $k_1,k_2 \in {\mathcal O}(n_0)$, while $(k_1+ q),(k_2-q) \in {\mathcal U}(m_0, n_0)$ [see Eqs.~(\ref{eq:O(n0)}) and (\ref{eq:U(m0,n0)}), respectively]; these conditions may be conveniently enforced by means of the CO-occupation function:
\begin{equation}\label{eq:n(k)}
n(k) = \left\{
\begin{array}{lcl}
          1&    &\mbox{for  } k \in  {\mathcal O}(n_0) \,, \\ 
          0&    &\mbox{for  } k \in  {\mathcal U}(m_0, n_0) \,,  
\end{array}
\right.
\end{equation}
and an auxiliary function
\begin{equation}\label{eq:overlinen(k)}
\overline{n}(k) = 1 - n(k) \,,
\end{equation}
where $k \in {\mathcal A}(m_0)$, see Eq.~(\ref{eq:A(m0)}). When the $t_2$ amplitudes $t(k_1, k_2, q)$ are derived from the exact eigenstate $\Psi_0$ (e.g., obtained by
using the FCI method), the {\em exact} electronic-correlation energy per molecule in our X$_M$ system, ${\mathcal E}_{\rm corr} \equiv {\mathcal E}_{\rm corr}^N$, may be
expressed as follows:
\begin{widetext}
\begin{equation}\label{eq:Ecorr}
{\mathcal E}_{\rm corr}  \equiv  E_{\rm corr}/M 
 =  M^{-1} \sum_{k_1,k_2,q \in {\mathcal A}(m_0)} 
t(k_1, k_2, q) 
\big[ 2 v(q) - v(k_2 - k_1 - q) \big] 
    n(k_1) n(k_2) \overline{n}(k_1 + q) \overline{n}(k_2 - q) \,,
\end{equation}
\end{widetext}
Amplitudes $t(k_1, k_2, q)$ are assumed to vanish identically if $n(k_1) n(k_2) \overline{n}(k_1 + q) \overline{n}(k_1 -q) = 0$. Because Eq.~(\ref{eq:E(2M-N)=EN(3)}) holds
both for the exact (FCI) and the Hartree-Fock energies, one concludes that
\begin{equation}\label{eq:E(2M-N)=EN(5)}
{\mathcal E}_{\rm corr}^{2M - N} = {\mathcal E}_{\rm corr}^N \,,
\end{equation}
which is also true for approximate correlation energies, calculated by means of a truncated coupled-cluster method or the perturbation theory.

   The (full) coupled-cluster (CC) equations comprise a set of coupled nonlinear equations, from which the exact $t_n$ amplitudes of the CC operator (\ref{eq:exp(T)}) may be calculated. In
practice, one often neglects the coupling between the amplitudes corresponding to lower excitations ($t_1$, $t_2$, $\ldots$), and those corresponding to higher excitations,
thus arriving at some truncated CC scheme (CCSD, CCSDT, CCSDTQ, \ldots).  In Sec.~\ref{sec:intro} we pointed out that attempts to find quantities $t(k_1, k_2, q)$ by using
the CCD method, or related methods which approximately include the coupling between the double and quadruple excitations (ACP, ACPQ), seem hopeless; also the CCDT and
CCDTQ methods are expected to fail in this respect.~\cite{podeszwa:02a} However, it turns out that one can calculate the perturbation-theory approximants to amplitudes
$t(k_1, k_2, q)$. To this end, Hamiltonian (\ref{eq:PPPH(2)}) has to be partitioned into the unperturbed (zero-order) part $\hat{H}^{(0)}$, being
the sum of the first two terms on the rhs of Eq.~(\ref{eq:PPPH(2)}), and the perturbation (first-order) operator $\hat{H}^{(1)}$, equal to the third term. 
One may now express the CC amplitudes as $t_n = t_n^{(1)} +
t_n^{(2)} + \cdots$, and solve the full set of CC equations order-by-order~\cite{monkhorst:81} by keeping the terms of the same order together (this perturbation approach
effectively linearize the CC equations). The perturbation series of the $t_2$-amplitude corrections, $t_2 = t_2^{(1)} + t_2^{(2)} + \cdots$, substituted into the correlation-energy
formula (\ref{eq:Ecorr}), defines the perturbation series of correlation-energy corrections ${\mathcal E}_{\rm corr} = {\mathcal E}_{\rm corr}^{(2)} + {\mathcal E}_{\rm corr}^{(3)}
+ \cdots$. 

\begin{figure}
\includegraphics[width=1.0\linewidth]{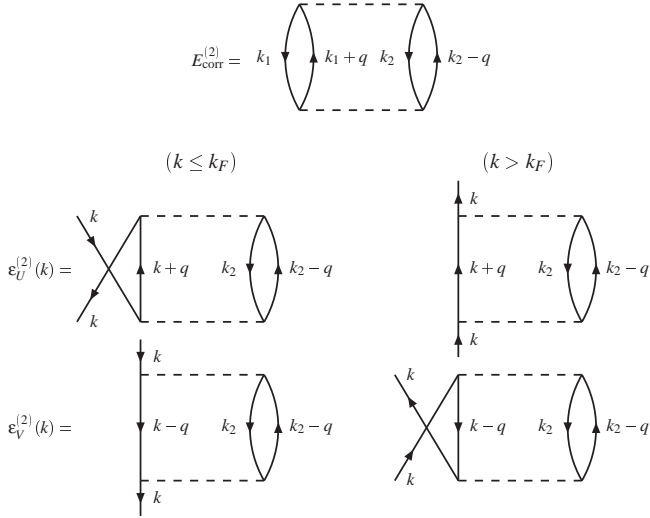}%
\caption{\label{fig:Brandows}
Brandow diagrams representing:  total second-order electronic-correlation energy 
$E_{\rm corr}^{(2)}$, and second-order contributions to electronic-energy bands, $\varepsilon_U^{(2)}(k)$ and $\varepsilon_V^{(2)}(k)$. Summation over spin indices is assumed.}
\end{figure}

   In the lowest order of the perturbation theory $\hat{T} = \hat{T}_2^{(1)}$, and the corresponding spin-adapted $t_2^{(1)}$ amplitudes of our 1D-metal model read
simply as
\begin{equation}\label{eq:t2(1)}
t^{(1)}(k_1, k_2, q) = - D^{-1}(k_1, k_2, q)  v(q)  \,,
\end{equation}
where 
\begin{widetext}
\begin{equation}\label{eq:D(k1,k2,q)}
D(k_1, k_2, q)  =  \varepsilon_{\rm HF}(k_1 + q) - \varepsilon_{\rm HF}(k_1)
+  \varepsilon_{\rm HF}(k_2 - q)  - \varepsilon_{\rm HF}(k_2)  
  =  D(k_1, k_2,k_2 - k_1 - q) = D(k_2, k_1, -q) \,,
\end{equation}
\end{widetext}
is an orbital-energy denominator. By substituting Eq.~(\ref{eq:t2(1)}) into formula (\ref{eq:Ecorr}) one obtains the expression for ${\mathcal E}_{\rm corr}^{(2)}$, the
second order correlation energy (per molecule) for our X$_M$ system 
The total second-order correlation energy, $E_{\rm corr}^{(2)} = M {\mathcal E}_{\rm corr}^{(2)}$, is represented by the topmost Brandow
diagram~\cite{brandow:67} depicted in Fig.~\ref{fig:Brandows}. It can be shown that $D(k_1, k_2, q) \ge 2 \Delta \varepsilon_{\rm HF} > 0$ [see Eq.~(\ref{eq:epsHFgap})], if
$n(k_1) n(k_2) \overline{n}(k_1 + q) \overline{n}(k_1 -q) \neq 0$, which makes ${\mathcal E}_{\rm corr}^{(2)}$ finite (and $< 0$) for finite~$M$.

\begin{figure}
\includegraphics{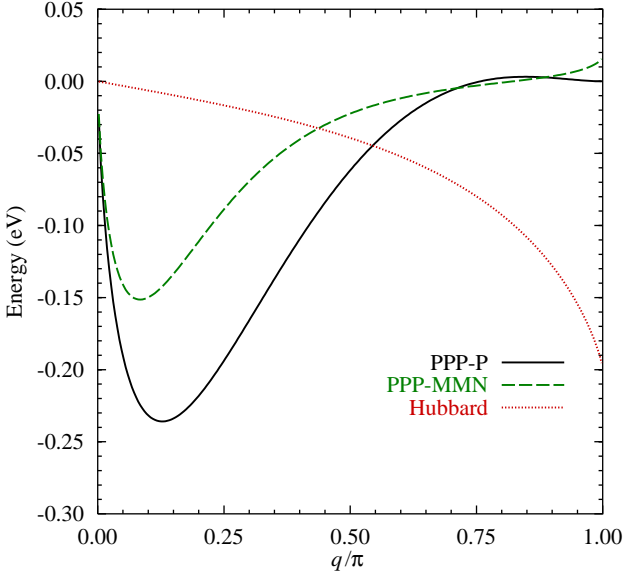}%
\caption{(Color online) \label{fig:F(q)pi/2}
$k_{\rm F} = \pi/2$. Functions $F^{(2)}(q)$ for PPP-P, PPP-MMN, and Hubbard-$0$ models.}
\end{figure}

\begin{figure}
\includegraphics{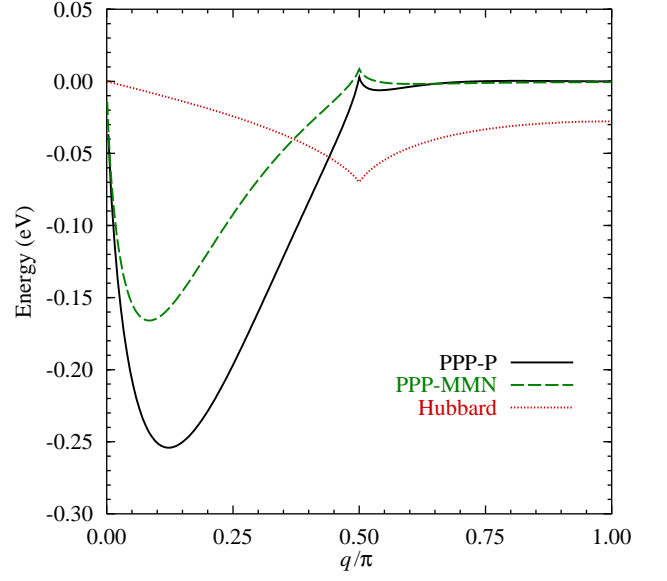}%
\caption{(Color online) \label{fig:F(q)pi/4}
$k_{\rm F} = \pi/4$. Functions $F^{(2)}(q)$ for PPP-P, PPP-MMN, and Hubbard-$0$ models.}
\end{figure}

  In order to study the extended system limit (\ref{eq:extsyslim}), we use the approach described in detail in the previous section. It is useful to write formula for
${\mathcal E}_{\rm corr}^{(2)}$ in the following form:
\begin{equation}\label{eq:E2corr}
{\mathcal E}_{\rm corr}^{(2)} = \int_0^{\pi} dq \, F^{(2)}(q) \,,
\end{equation}
where, for $q \in [0, \pi]$,
\begin{align}\label{eq:F2(q)}
F^{(2)}(q) &=  - \frac{V(q)}{4 \pi^3} \int_{-k_{\rm F}}^{k_{\rm F}} dk_1 
\int_{-k_{\rm F}}^{k_{\rm F}} dk_2 \,
D^{-1}(k_1, k_2, q) \nonumber\\
&\quad{}\times\big[ 2 V(q) - V(k_2 - k_1 - q) \big] 
\overline{n}(k_1 + q) \overline{n}(k_2 - q) \,,
\end{align}
and for the definition of function $V(q)$, see Eqs.~(\ref{eq:V(q)}) and (\ref{eq:VP(q)(2)}).  Function $\overline{n}(k)$, corresponding to Eq.~(\ref{eq:overlinen(k)}),
becomes now the characteristic function for the union of intervals $(-\pi, -k_{\rm F})$ and $(k_{\rm F}, \pi]$. It is not certain {\it a priori\/} whether quantity
(\ref{eq:E2corr}) is finite, since the definition of function $F^{(2)}(q)$ in Eq.~(\ref{eq:F2(q)}) involves singular quantities. Let us note that the analysis by Beleznay {\it et al.\/},~\cite{beleznay:81} employed by Suhai and Ladik,~\cite{suhai:82} does not apply to the case of the PPP model, since, for $q \rightarrow 0$, neither the denominator (\ref{eq:D(k1,k2,q)}) is linear in $q$, nor the numerator involving function $V(q)$ remains finite in this limit. (However, the analysis of Ref.~\onlinecite{beleznay:81} is valid for the Hubbard-$n$ models.) By analytical and numerical means, we have been able to show that, for the PPP model of 1D metal, function $F^{(2)}(q)$ is finite, and thus leads to a finite value of ${\mathcal E}_{\rm corr}^{(2)}$. In particular, for $q \rightarrow 0$ the asymptotic form of function $F^{(2)}(q)$ reads as
\begin{equation}\label{eq:F2PPPasympt}
F^{(2)}(q) \rightarrow \frac{e^2}{\pi^2 R^0} \, q \ln q \,,
\end{equation}
for the PPP model (independently of the form of the $\gamma$-potential), and 
\begin{equation}\label{eq:F2Hubasympt}
F^{(2)}(q) \rightarrow \frac{(\gamma^0)^2}{16 \pi^3 \beta^0 \sin k_{\rm F}} \; q \,,
\end{equation}
for the Hubbard-$0$ model. Functions $F^{(2)}(q)$ corresponding to the PPP-P, PPP-MMN, and Hubbard-$0$ models, for $k_{\rm F } = \pi/2$ and $\pi/4$, are presented in
Figs.~\ref{fig:F(q)pi/2} and \ref{fig:F(q)pi/4}, respectively. These results were obtained by applying a numerical-integration routine based on Romberg's
algorithm.~\cite{numrecC:92} The behavior of these functions in the vicinity of $q = 0$ agrees with the predictions of Eqs.~(\ref{eq:F2PPPasympt}) and
(\ref{eq:F2Hubasympt}). In addition, each function develops a cusp at $q = 2 k_{\rm F }$ [the cusp is absent for the PPP-P model with $k_{\rm F } = \pi/2$, because in this
case $V_{\rm P}(\pi) = 0$].

   At the qualitative level, functions $F^{(2)}(q)$ corresponding to the PPP-P and PPP-MMN models of the 1D metal are similar, but that of the Hubbard-$0$ model is
distinctly different. On the other hand, the plot of $F^{(2)}(q)$ for the PPP model exhibits a striking similarity to the plot of an analogous function, corresponding to the CC method, calculated for the 3D electron-gas model by Bishop and L\"{u}hrmann,~\cite{bishop:78,bishop:82} and by Emrich and Zabolitzky.~\cite{emrich:84} This similarity points to some common behavior of electronic correlations in 3D and 1D metallic systems, which in the latter case, at least qualitatively, may be described within the second-order perturbation theory.

\begin{figure}
\includegraphics{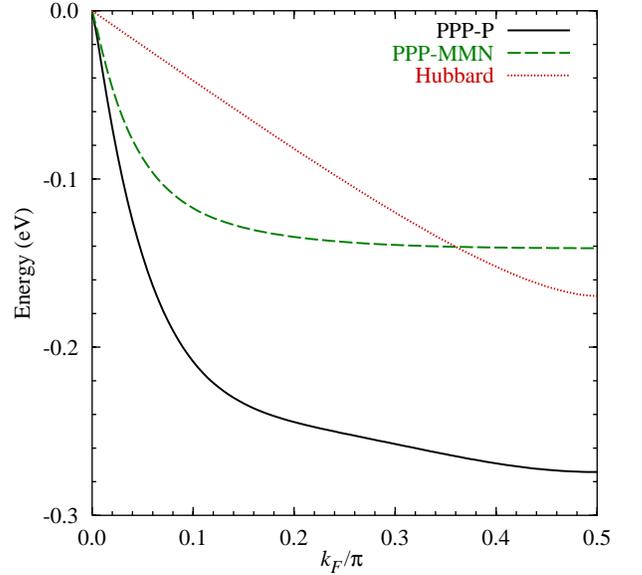}%
\caption{(Color online) \label{fig:E2(kF)}
Second-order electronic-correlation energy per molecule, ${\mathcal E}_{\rm corr}^{(2)}$, as a function of $k_{\rm F }/\pi$ for PPP-P, PPP-MMN, and Hubbard-$0$ models.}
\end{figure}

   The second-order electronic-correlation energy per molecule (\ref{eq:E2corr}) is a function of the Fermi momentum, ${\mathcal E}_{\rm corr}^{(2)} \equiv {\mathcal E}_{\rm
corr}^{(2) k_{\rm F }}$, and fulfills condition
\begin{equation}\label{eq:E(p-kF)=EkF(2)}
{\mathcal E}_{\rm corr}^{(2) (\pi - k_{\rm F })} = {\mathcal E}_{\rm corr}^{(2) k_{\rm F }} \,,
\end{equation}
c.f. Eq.~(\ref{eq:E(2M-N)=EN(5)}). In Fig.~\ref{fig:E2(kF)} we draw ${\mathcal E}_{\rm corr}^{(2)k_{\rm F }}$, corresponding to the PPP-P, PPP-MMN, and Hubbard-$0$ models, for $ k_{\rm F }/\pi \in [0,1/2]$. By comparing the PPP-P and PPP-MMN models at the second-order perturbation theory and the Hartree-Fock levels (for the latter, see Fig.~\ref{fig:EHF(kF)}), one may conclude that the second-order electronic-correlation corrections are more sensitive to the form of the $\gamma$ potential than are the Hartree-Fock results. Our second-order results for the Hubbard-$0$ model agree with those of Metzner and Vollhardt.~\cite{metzner:89} 

The convergence pattern of the finite-system PPP-P results toward the extended system limit is studied in Table~\ref{tab:EHF,E2}.
The extended-system Hartree-Fock results are obtained from formulas (\ref{eq:EHF(3a)}) and (\ref{eq:EHF(3b)}). It is seen that our numerical-integration scheme applied to the calculation of the second-order energies in the extended-system limit [by using Eqs.~(\ref{eq:E2corr}) and (\ref{eq:F2(q)})] gives very good results (although getting high-accuracy results required intense computations using a very fine grid). The data in Table~\ref{tab:EHF,E2} indicate that the errors in the Hartree-Fock and the electronic-correlation energies, originating from the finite-size effects, partially cancel out.

\begin{table*}
\caption{\label{tab:EHF,E2}
PPP-P model. Hartree-Fock energies (HF) and second-order correlation energies (MP2)  per molecule (in eV) for X$_M$ system, shown for extended-system limit ($M = \infty$) and for finite $M$ (as energy
differences with respect to $M = \infty$).}

\begin{ruledtabular}
\begin{tabular}{crrrrrrrr}

\multicolumn{9}{c}{$k_{\rm F}=\pi/4$}\\
\hline
         $M$ & \multicolumn{1}{c}{$\infty$}& \multicolumn{1}{c}{$   15996$} & \multicolumn{1}{c}{$    7996$} & \multicolumn{1}{c}{$    3996$} & \multicolumn{1}{c}{$    1996$} & \multicolumn{1}{c}{$     996$} & \multicolumn{1}{c}{$     100$} & \multicolumn{1}{c}{$      20$} \\
\hline	 
HF & $-2.729\mbox{ }774$ & $-0.000\mbox{ }000$ & $-0.000\mbox{ }001$ & $-0.000\mbox{ }002$ & $-0.000\mbox{ }007$ & $-0.000\mbox{ }027$ & $-0.001\mbox{ }926$ & $-0.034\mbox{ }256$ \\
MP2 & $-0.251\mbox{ }402$ & $0.000\mbox{ }000$ & $0.000\mbox{ }002$ & $0.000\mbox{ }005$ & $0.000\mbox{ }018$ & $0.000\mbox{ }057$ & $0.002\mbox{ }188$ & $0.021\mbox{ }514$ \\
\hline
\multicolumn{9}{c}{$k_{\rm F}=\pi/2$}\\
\hline
         $M$ & \multicolumn{1}{c}{$\infty$}& \multicolumn{1}{c}{$   15998$} & \multicolumn{1}{c}{$    7998$} & \multicolumn{1}{c}{$    3998$} & \multicolumn{1}{c}{$    1998$} & \multicolumn{1}{c}{$     998$} & \multicolumn{1}{c}{$     102$} & \multicolumn{1}{c}{$      22$} \\
\hline	 
HF & $-1.810\mbox{ }654$ & $-0.000\mbox{ }000$ & $-0.000\mbox{ }001$ & $-0.000\mbox{ }002$ & $-0.000\mbox{ }008$ & $-0.000\mbox{ }030$ & $-0.002\mbox{ }119$ & $-0.034\mbox{ }663$ \\
MP2 & $-0.274\mbox{ }249$ & $0.000\mbox{ }000$ & $0.000\mbox{ }001$ & $0.000\mbox{ }005$ & $0.000\mbox{ }016$ & $0.000\mbox{ }050$ & $0.001\mbox{ }903$ & $0.017\mbox{ }587$ \\ 
\end{tabular}
\end{ruledtabular}
\end{table*}

The 1D-metal model introduced in the present paper may be used to generate benchmark results for testing the {\it ab initio} computer programs for conducting polymers. In his study of a chain of equidistant hydrogen atoms, Suhai~\cite{suhai:94a} applied a cut-off radius of $16$~\AA\/ in the calculations of the exchange and correlation terms, which should correspond to our results for $M = 20, 22$ in Table~\ref{tab:EHF,E2}. Thus, one may expect that the accuracy of the results of Ref.~\onlinecite{suhai:94a} is no worse than $0.05$~eV.

\section{ \label{sec:E2band}%
Second-order correlation corrections to electronic-energy bands of 1D-metal model}

  Theoretical models describing the influence of electronic-correlation effects on the electronic-energy bands in solids were studied by Toyozawa,~\cite{toyozawa:54}
Kunz,~\cite{kunz:72} and Pantelides {\it et al.\/}~\cite{pantelides:74} The departure point for these models was the second-order perturbation 
theory of correlation effects, applied to the
ground state, and to some ionized states of the solid. Recently, Sun and Bartlett~\cite{sun:96a,sun:96b,sun:97,sun:98,sun:99} developed an {\it ab initio} perturbation approach to
the band structure of solids with a non-zero band gap. Below we shall analyze the correlation corrections to the Hartree-Fock energy bands of our 1D-metal model and perform
calculations of these corrections within the second-order perturbation theory.

  A formal definition of the ``exact single-electron energies,'' $\varepsilon(k)$, in  the X$_M$ system may be modeled after Koopmans' theorem:~\cite{koopmans:34} 
\begin{equation}\label{eq:eps(k)}
\varepsilon(k) = \left\{ 
\begin{array}{rl}
E_0^N - E_{k, \sigma}^{N - 1} \,, & \mbox{  for  }  k \in {\mathcal O}(n_0) \,, \\
E_{k, \sigma}^{N + 1} - E_0^N \,, & \mbox{  for  }  k \in {\mathcal U}(m_0,n_0) \,.
\end{array}
\right.
\end{equation}
In the above formula, $\varepsilon(k) \equiv \varepsilon^N(k)$ is expressed through the differences of certain eigenenergies of Hamiltonian~(\ref{eq:PPPH(1)}): as before,
$E_0^N$ corresponds to the $N$-electron closed-shell ground state $\Psi_0^N$, while $E_k^{N - 1}$, and $E_k^{N + 1}$ correspond to some spin-doublet eigenstates, 
$\Psi_{k,\sigma}^{N - 1}$ and $\Psi_{k, \sigma}^{N + 1}$, where the lower index indicates the $k$th representation of the translation group ${\mathcal T}_M$, cf.
Eq.~(\ref{eq:COs}), and $\sigma$ is a spin index ($\sigma = \alpha$ or $\beta$). It is assumed that functions $\Psi_{k, \sigma}^{N - 1}$ and $\Psi_{k, \sigma}^{N + 1}$ are 
similar (in the sense of maximum overlap) to the determinantal functions $\Phi_{k, \sigma}^{N - 1}$ and $\Phi_{k, \sigma}^{N + 1}$, respectively. Function $\Phi_{k,
\sigma}^{N - 1}$ is obtained from the restricted Hartree-Fock (RHF) ground-state function $\Phi_{\rm HF} \equiv \Phi_0^N$ by removing 
spinorbital $\psi_k \sigma$; in turn, $\Psi_{k, \sigma}^{N + 1}$ is obtained from $\Phi_{\rm HF}$ by adding spinorbital $\psi_k \sigma$.  Due to the high symmetry of our model, functions $\Phi_{k, \sigma}^{N - 1}$ and 
$\Phi_{k,\sigma}^{N + 1}$ appear to be the open-shell RHF wave functions for the respective doublet states. Let us note that the above definition of functions 
$\Psi_{k, \sigma}^{N- 1}$ and $\Psi_{k, \sigma}^{N + 1}$ is, in fact, an oversimplification, but it is sufficient for a theory involving only the second-order corrections.~\cite{toyozawa:54,kunz:72,pantelides:74} 
The physical sense of quantities defined in Eq.~(\ref{eq:eps(k)}) is as follows: for $k \in {\mathcal O}(n_0)$, $-\varepsilon(k)$ represent certain ionization potentials of the X$_M$ system, 
while for $k \in {\mathcal U}(m_0,n_0)$, $\varepsilon(k)$ are equal to some electron affinities of this system. Due to the time-reversal symmetry, $\varepsilon(k) = 
\varepsilon(-k)$.

   By partitioning the eigenenergies into the Hartree-Fock and correlation parts, one may define the electronic-correlation contribution to $\varepsilon(k)$:
\begin{equation}\label{eq:epscorr(k)}
\varepsilon_{\rm corr}(k) = \varepsilon(k) - \varepsilon_{\rm HF}(k) \,,
\end{equation}
and $\varepsilon_{\rm corr}(k) \equiv \varepsilon_{\rm corr}^N(k)$ may be calculated as the difference of the pertinent correlation energies. Due to the alternancy symmetry of
Hamiltonian~(\ref{eq:PPPH(1)}), the differences of eigenenergies in Eq.~(\ref{eq:eps(k)}) 
obey formula~(\ref{eq:E(2M-N)=EN(4)}), which leads to a relation for
$\varepsilon(k)$ analogous to that for the CO energies in Eq.~(\ref{eq:altsymepsHF}). As a consequence, the correlation contributions~(\ref{eq:epscorr(k)}) fulfill the
following condition:
\begin{equation}\label{eq:altsymepscorr}
\varepsilon_{\rm corr}^{2M-N}(k) = -\varepsilon_{\rm corr}^N(k - q_0) \,,
\end{equation}
where $q_0 = 2m_0 +1$. 
  
  By calculating the second-order correlation energies corresponding to the states involved in definitions~(\ref{eq:eps(k)}) and~(\ref{eq:epscorr(k)}), one may obtain the second-order
correlation corrections to the energy bands, $\varepsilon_{\rm corr}^{(2)}(k) \equiv \varepsilon_{\rm corr}^{(2) N}(k)$. These correlation corrections emerge as differences of
extensive quantities (the total correlation energies of the respective states), but due to the exact cancellation of certain extensive contributions, the resulting formula
corresponds to an intensive quantity. By using the spin-adapted formulas of Ref.~\onlinecite{sun:96a}, one may write
\begin{equation}\label{eq:eps2(k)}
\varepsilon_{\rm corr}^{(2)}(k) =  \varepsilon_{U}^{(2)}(k) + \varepsilon_{V}^{(2)}(k) \,,
\end{equation}
where
\begin{widetext}
\begin{equation}\label{eq:epsU(k)}
\varepsilon_{U}^{(2)}(k) =  - \sum_{k_2,q \in {\mathcal A}(m_0)} D^{-1}(k, k_2, q)
v(q) \big[ 2 v(q) - v(k_2 - k - q) \big] 
n(k_2) \overline{n}(k + q) \overline{n}(k_2 - q) \,,
\end{equation}
and    
\begin{equation}\label{eq:epsV(k)}
\varepsilon_{V}^{(2)}(k) =  \sum_{k_2,q \in {\mathcal A}(m_0)} D^{-1}(k - q, k_2, q)
v(q) \big[ 2 v(q) - v(k_2 - k) \big] 
n(k_2) n(k - q) \overline{n}(k_2 - q) \,.
\end{equation}
\end{widetext}
In the above formulas we use definitions~(\ref{eq:n(k)}), (\ref{eq:overlinen(k)}), and (\ref{eq:D(k1,k2,q)}). The Brandow diagrams representing the second-order contributions 
$\varepsilon_{U}^{(2)}(k) \equiv \varepsilon_{U}^{(2) N}(k)$ and 
$\varepsilon_{V}^{(2)}(k) \equiv \varepsilon_{V}^{(2) N}(k)$ are depicted in Fig.~\ref{fig:Brandows}. These quantities may be related to the second-order ground-state correlation energy:
\begin{align}\label{eq:E2corrUV}
{\mathcal E}_{\rm corr}^{(2)} &=
M^{-1} \sum_{k \in {\mathcal A}(m_0)} \varepsilon_{U}^{(2)}(k) \, n(k) \nonumber\\
&=  - M^{-1} \sum_{k \in {\mathcal A}(m_0)} \varepsilon_{V}^{(2)}(k) \, \overline{n}(k) \,,
\end{align}
cf. Eqs.~(\ref{eq:Ecorr}) and (\ref{eq:t2(1)}). It can be shown that, in addition to relation~(\ref{eq:altsymepscorr}) holding for $\varepsilon_{\rm corr}^{(2)}(k)$, one has also
\begin{equation}\label{eq:altsymepsUV}
\varepsilon_{V}^{(2) \, 2M-N}(k) = -\varepsilon_{U}^{(2) \, N}(k - q_0) \,.
\end{equation}

   In the extended-system limit~(\ref{eq:extsyslim}), the correlation corrections to the energy bands for our 1D-metal model, $\varepsilon_{\rm corr} \equiv \varepsilon_{\rm
corr}^{k_{\rm F}}(k)$, fulfill the analogue of relation ~(\ref{eq:altsymepscorr}):
\begin{equation}\label{eq:altsymepscorr(2)}
\varepsilon_{\rm corr}^{\pi - k_{\rm F}}(k) = -\varepsilon_{\rm corr}^{k_{\rm F}}(k - \pi) \,.
\end{equation}
 For $k_{\rm F} = \pi/2$ (the half-filling case) the above formula implies that the {\em exact\/} correlation correction must vanish at the Fermi level:
\begin{equation}\label{eq:altsymepscorr(3)}
\varepsilon_{\rm corr}^{\pi/2}(\pm \pi/2) = 0 \,.
\end{equation}
In the extended-system limit, by using the approach described in Sec.~\ref{sec:HF}, one may rewrite the second-order correlation corrections~(\ref{eq:epsU(k)}) and~(\ref{eq:epsV(k)}) as
\begin{align}\label{eq:epsU(k)(2)}
\varepsilon_{U}^{(2)}(k) & = - \frac{1}{4 \pi^2} \int_{-\pi}^{\pi} dq \,  V(q) 
\int_{-k_{\rm F}}^{k_{\rm F}} dk_2 \, D^{-1}(k, k_2, q)
\nonumber \\
& \quad{} \times\big[ 2 V(q) - V(k_2 - k - q) \big]
 \overline{n}(k + q) \overline{n}(k_2 - q) \,,
\end{align}
and    
\begin{align}\label{eq:epsV(k)(2)}
\varepsilon_{V}^{(2)}(k) & =  \frac{1}{4 \pi^2} \int_{-\pi}^{\pi} dq \,  V(q) 
\int_{-k_{\rm F}}^{k_{\rm F}} dk_2 \, D^{-1}(k - q, k_2, q) 
\nonumber \\ 
& \quad{} \times
\big[ 2 V(q) - V(k_2 - k) \big]
 n(k - q) \overline{n}(k_2 - q) \,,
\end{align}
where now $\varepsilon_{U}^{(2)}(k) \equiv \varepsilon_{U}^{(2) k_{\rm F}}(k)$ and $\varepsilon_{V}^{(2)}(k) \equiv \varepsilon_{V}^{(2) k_{\rm F}}(k)$. The full second-order energy-band
corrections~(\ref{eq:eps2(k)}) fulfill formulas~(\ref{eq:altsymepscorr(2)}) and~(\ref{eq:altsymepscorr(3)}); in addition, the analogue of formula~(\ref{eq:altsymepsUV})
reads as
\begin{equation}\label{eq:altsymepsUV(2)}
\varepsilon_{V}^{(2) \,\pi - k_{\rm F}}(k) = -\varepsilon_{U}^{(2) \, k_{\rm F}}(k - \pi) \,,
\end{equation}
which, for $k_{\rm F} = \pi/2$, gives 
\begin{equation}\label{eq:altsymepsUV(3)}
\varepsilon_{V}^{(2) \, \pi/2}(\pm \pi/2) = -\varepsilon_{U}^{(2) \, \pi/2}(\pm \pi/2) \,,
\end{equation}
in agreement with the general condition~(\ref{eq:altsymepscorr(3)}).

  We have studied the behavior of the second-order correlation corrections to the energy bands of the X$_M$ system within the PPP-P and Hubbard-$0$ models. As indicated in subsection~\ref{ssec:EssparPPPH}, the former model is a convenient, single-parameter reference PPP model for 1D metals. Let us note that the PPP-P model shares basic qualitative features of other PPP models, perhaps with some amplification of the electronic-correlation effects, see Fig.~\ref{fig:E2(kF)}. 

  In our energy-band calculations we found that already for finite $M$ some numerical problems emerged when $N \neq M$. For instance, in Eq.~(\ref{eq:epsU(k)}) denominator $D(k,k_2,q)$ may change its sign when going from one
allowed triple of indices $k,k_2,q \in {\mathcal A}(m_0)$ to another. Moreover, in some cases the corresponding denominator may closely approach zero, thus leading to an
(almost) divergent result. Similar problems emerge in Eq.~(\ref{eq:epsV(k)}) for denominators $D(k-q,k_2,q)$. In the case of the PPP-P model, such problems affect
$\varepsilon_{U}^{(2)}(k)$ for $N < M$ and $k \in {\mathcal U}(m_0,n_0)$; by the virtue of Eq.~(\ref{eq:altsymepsUV}), the same problems affect $\varepsilon_{V}^{(2)}(k)$ for $N
> M$ and $k \in {\mathcal O}(n_0)$. We thus conclude that in the extended-system limit of the PPP model the second-order correlation corrections $\varepsilon_{\rm corr}^{(2)}(k) =
\varepsilon_{\rm corr}^{(2) k_{\rm F}}(k)$ are not defined for $|k| \ge k_{\rm F}$, if $k_{\rm F} < \pi/2$, and for $|k| \le k_{\rm F}$, if $k_{\rm F} > \pi/2$. Only for
$k_{\rm F} = \pi/2$ the second-order correlation corrections (\ref{eq:eps2(k)}) are defined for all values of $k$.

\begin{figure}
\includegraphics{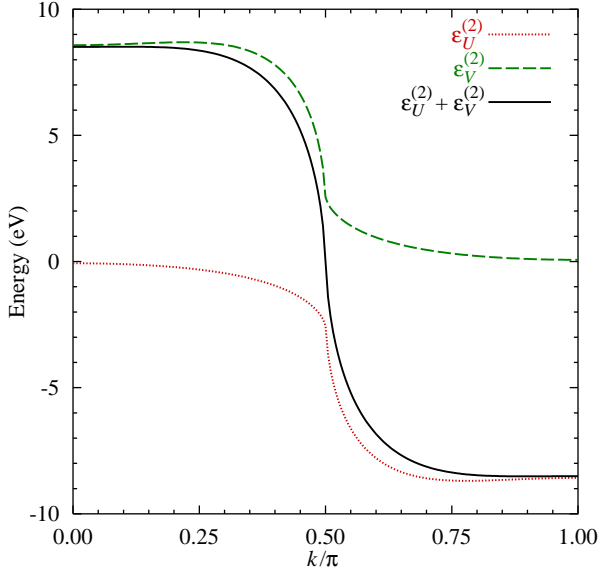}%
\caption{(Color online) \label{fig:PPP-Peps2(k)pi/2}
PPP-P model, $k_{\rm F} = \pi/2$. Second-order corrections to electronic-energy bands, $\varepsilon_{U}^{(2)}(k)$ and $\varepsilon_{V}^{(2)}(k)$, and their sum 
[$= \varepsilon_{\rm corr}^{(2)}(k)$], shown as functions of $k/\pi$.}
\end{figure}

\begin{figure}
\includegraphics{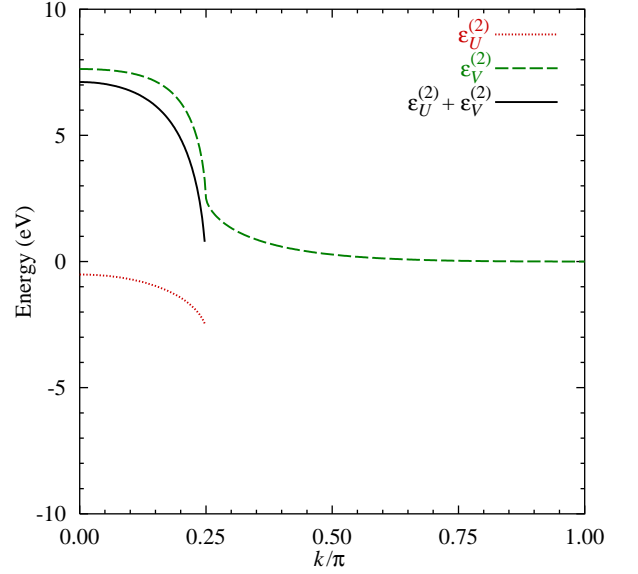}%
\caption{(Color online) \label{fig:PPP-Peps2(k)pi/4}
PPP-P model, $k_{\rm F} = \pi/4$. Second-order corrections to electronic-energy bands, $\varepsilon_{U}^{(2)}(k)$ and $\varepsilon_{V}^{(2)}(k)$, and their sum 
[$= \varepsilon_{\rm corr}^{(2)}(k)$], shown as functions of $k/\pi$.}
\end{figure}

For the PPP-P model, quantities $\varepsilon_{U}^{(2)}(k)$ and $\varepsilon_{V}^{(2)}(k)$, and their sum [$= \varepsilon_{\rm corr}^{(2)}(k)$], are shown in
Fig.~\ref{fig:PPP-Peps2(k)pi/2} (for $k_{\rm F} = \pi/2$), and in Fig.~\ref{fig:PPP-Peps2(k)pi/4} (for $k_{\rm F} = \pi/4$). It is seen that corrections
$\varepsilon_{U}^{(2)}(k)$ and $\varepsilon_{V}^{(2)}(k)$ are continuous functions of quasimomentum $k$ [except that $\varepsilon_{U}^{(2) \pi/4}(k)$ is not defined 
for $|k| >\pi/4$], but, apparently, are not smooth for $k = \pm k_{\rm F}$. The full second-order correlation correction, $\varepsilon_{\rm corr}^{(2)}(k)$, is continuous, but has a diverging derivative at
$k = \pm k_{\rm F}$. In the PPP-P model, one finds that $\varepsilon_{U}^{(2)}(k) < 0$, while $\varepsilon_{V}^{(2)}(k) > 0$, for those values of $k$, for which these quantities can be calculated.

\begin{figure}
\includegraphics{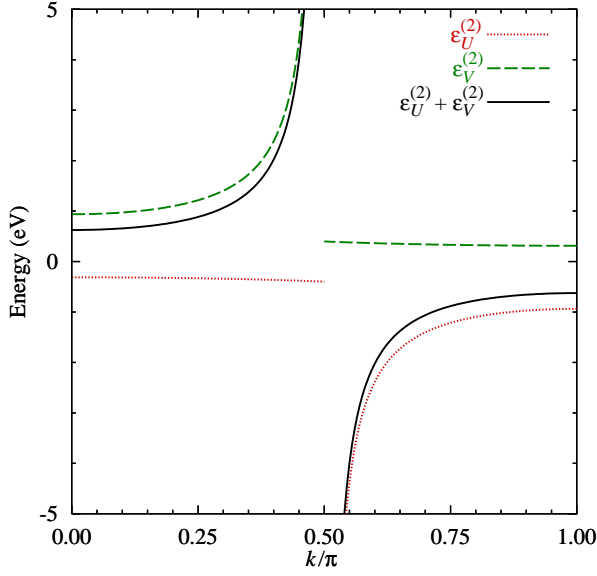}%
\caption{(Color online) \label{fig:Hubeps2(k)pi/2}
Hubbard-$0$ model, $k_{\rm F} = \pi/2$. Second-order corrections to electronic-energy bands, $\varepsilon_{U}^{(2)}(k)$ and $\varepsilon_{V}^{(2)}(k)$, and their sum 
[$= \varepsilon_{\rm corr}^{(2)}(k)$], shown as functions of $k/\pi$.}
\end{figure}

\begin{figure}
\includegraphics{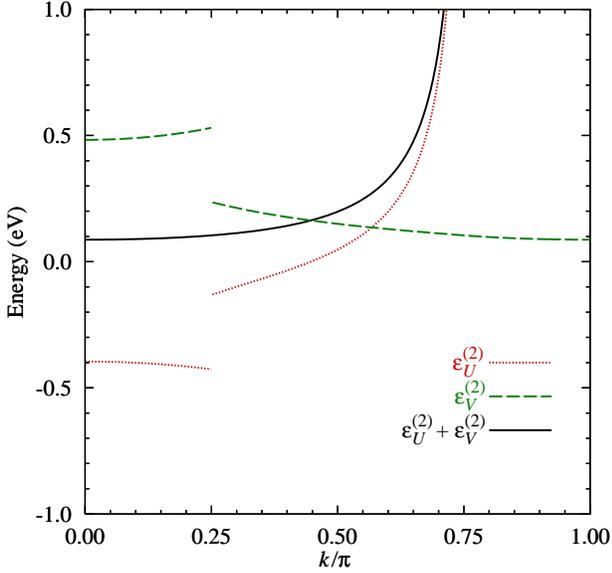}%
\caption{(Color online) \label{fig:Hubeps2(k)pi/4}
Hubbard-$0$ model, $k_{\rm F} = \pi/4$. Second-order corrections to electronic-energy bands, $\varepsilon_{U}^{(2)}(k)$ and $\varepsilon_{V}^{(2)}(k)$, and their sum 
[$= \varepsilon_{\rm corr}^{(2)}(k)$], shown as functions of $k/\pi$.}
\end{figure}

For the Hubbard-$0$ model, a different behavior of the correlation corrections is found, see Fig.~\ref{fig:Hubeps2(k)pi/2} (for $k_{\rm F} = \pi/2$), and
Fig.~\ref{fig:Hubeps2(k)pi/4} (for $k_{\rm F} = \pi/4$). A special case correspond to the half-filled band: $\varepsilon_{U}^{(2)}(k)$ and $\varepsilon_{V}^{(2)}(k)$ are defined everywhere except for $k = \pm \pi/2$, where they diverge. For $k_{\rm F} \neq \pi/2$ the problems with the denominators, similar to those discussed above for the PPP-P model,
emerge:  if $k_{\rm F} < \pi/2$, quantity $\varepsilon_{U}^{(2)}(k)$ is not defined for $|k| \ge \pi - k_{\rm F}$; in turn, if $k_{\rm F} > \pi/2$, quantity
$\varepsilon_{V}^{(2)}(k)$ is not defined for $|k| \le \pi - k_{\rm F}$. In both cases, the corresponding quantity diverges when approaching the forbidden region.
Interestingly, for $k_{\rm F} \neq \pi/2$ both $\varepsilon_{U}^{(2)}(k)$ and $\varepsilon_{V}^{(2)}(k)$ are discontinuous at $k = \pm k_{\rm F}$
[the gaps amount to $\gamma^2_0/(48\beta^{0} \cos k_{\rm F})$], but their sum, $\varepsilon_{\rm
corr}^{(2)}(k)$, remains a continuous function at this point. As discussed in a review by Mila and Penc,~\cite{mila:01} according to the exact results of Lieb and
Wu,~\cite{lieb:68} the Hubbard-$0$ model of the 1D metal behaves as an insulator at the half filling, and as the so-called Luttinger liquid for $k_{\rm F} \neq \pi/2$. The
divergence of $\varepsilon_{\rm corr}^{(2)}(k)$ at $k = \pm \pi/2$ for $k_{\rm F} = \pi/2$ is likely to reflect the inadequacy of the band-structure picture for the ground
state of the Hubbard-$0$ model at the half filling.

For the PPP-P model, we also studied the convergence of the finite-system results toward the extended-system limit. In Tables~\ref{tab:epsHF,eps2,pi/2} and~\ref{tab:epsHF,eps2,pi/4} we show the convergence of quantities $\varepsilon_{\rm HF}(k)$, $\varepsilon_{U}^{(2)}(k)$ and $\varepsilon_{V}^{(2)}(k)$ for $k_{\rm F} = \pi/2$ and $k_{\rm F} = \pi/4$, respectively. The results are presented for $k =0$, $k \approx k_{\rm F}/2$, and $k \approx k_{\rm F}$. It is found that for $\varepsilon_{V}^{(2)}(k)$ the finite-size effects decay extremely slow, especially for $k$ in the vicinity of $k_{\rm F}$, while $\varepsilon_{U}^{(2)}(k)$ and $\varepsilon_{\rm HF}(k)$ behave satisfactorily. In the extended system limit, the calculation of $\varepsilon_{V}^{(2)}(k)$ from formula~(\ref{eq:epsV(k)(2)}) also meets with numerical difficulties, and attaining the accuracy of $4$--$5$ decimal places by using of our Romberg-integration approach requires extremely fine grids. The results displayed in Table~\ref{tab:EHF,E2} indicate that the electronic-correlation corrections to the total energy (per molecule) in the 1D metal saturate rather quickly with the size of the system, and thus may be studied by employing a finite-system variant of the PPP model, as in Refs.~\onlinecite{fano:99} and \onlinecite{fano:98}. On the other hand, the results in Tables~\ref{tab:epsHF,eps2,pi/2} and~\ref{tab:epsHF,eps2,pi/4} do not encourage finite-system extrapolations in the case of the electronic-correlation corrections to the energy bands.

\begin{table}
\caption{\label{tab:epsHF,eps2,pi/2}
PPP-P model, $k_{\rm F} = \pi/2$. Hartree-Fock orbital energies and corresponding second-order correlation corrections (in eV) for  X$_M$ system, shown for extended-system limit ($M = \infty$), and for finite $M$ (as energy differences with respect to $M = \infty$).}
\begin{ruledtabular}
\begin{tabular}{rrrrrrr}
\multicolumn{7}{c}{$k=0$}\\
\hline
         $M$ & \multicolumn{1}{c}{$\infty$}& \multicolumn{1}{c}{$   40338$} & \multicolumn{1}{c}{$   13446$} & \multicolumn{1}{c}{$    4482$} & \multicolumn{1}{c}{$    1494$} & \multicolumn{1}{c}{$     498$} \\
\hline	 
$\varepsilon_{\rm HF}$ & $ -3.8683$ & $ -0.0000$ & $ -0.0000$ & $ -0.0000$ & $ -0.0000$ & $ -0.0000$ \\
$\varepsilon_U^{(2)}$ & $ -0.0687$ & $  0.0000$ & $  0.0000$ & $  0.0000$ & $  0.0000$ & $  0.0000$ \\
$\varepsilon_V^{(2)}$ & $  8.5785$ & $ -0.0092$ & $ -0.0244$ & $ -0.0633$ & $ -0.1605$ & $ -0.3946$ \\
\hline
\multicolumn{7}{c}{$k= 62\pi/249$}\\
\hline
         $M$ & \multicolumn{1}{c}{$\infty$}& \multicolumn{1}{c}{$   40338$} & \multicolumn{1}{c}{$   13446$} & \multicolumn{1}{c}{$    4482$} & \multicolumn{1}{c}{$    1494$} & \multicolumn{1}{c}{$     498$} \\
\hline	 
$\varepsilon_{\rm HF}$ & $ -1.3562$ & $ -0.0000$ & $ -0.0000$ & $ -0.0000$ & $ -0.0000$ & $ -0.0000$ \\
$\varepsilon_U^{(2)}$ & $ -0.3175$ & $  0.0000$ & $  0.0000$ & $  0.0000$ & $  0.0000$ & $  0.0000$ \\
$\varepsilon_V^{(2)}$ & $  8.6845$ & $ -0.0093$ & $ -0.0246$ & $ -0.0641$ & $ -0.1636$ & $ -0.4056$ \\
\hline
\multicolumn{7}{c}{$k=124\pi/249$}\\
\hline
         $M$ & \multicolumn{1}{c}{$\infty$}& \multicolumn{1}{c}{$   40338$} & \multicolumn{1}{c}{$   13446$} & \multicolumn{1}{c}{$    4482$} & \multicolumn{1}{c}{$    1494$} & \multicolumn{1}{c}{$     498$} \\
\hline	 
$\varepsilon_{\rm HF}$ & $  6.9583$ & $ -0.0000$ & $ -0.0000$ & $ -0.0000$ & $ -0.0006$ & $ -0.0006$ \\
$\varepsilon_U^{(2)}$ & $ -2.4095$ & $  0.0000$ & $  0.0000$ & $  0.0002$ & $  0.0020$ & $  0.0045$ \\
$\varepsilon_V^{(2)}$ & $  3.1720$ & $ -0.0136$ & $ -0.0406$ & $ -0.1177$ & $ -0.2884$ & $ -0.6819$ \\
\end{tabular}
\end{ruledtabular}
\end{table}

\begin{table}
\caption{\label{tab:epsHF,eps2,pi/4}
PPP-P model, $k_{\rm F} = \pi/4$. Hartree-Fock orbital energies and corresponding second-order correlation corrections (in eV) for X$_M$ system, shown for extended-system limit ($M =
\infty$), and for finite $M$ (as energy differences with respect to $M = \infty$).}
\begin{ruledtabular}
\begin{tabular}{rrrrrrr}
\multicolumn{7}{c}{$k=0$}\\
\hline
         $M$ & \multicolumn{1}{c}{$\infty$}& \multicolumn{1}{c}{$   80676$} & \multicolumn{1}{c}{$   26892$} & \multicolumn{1}{c}{$    8964$} & \multicolumn{1}{c}{$    2988$} & \multicolumn{1}{c}{$     996$} \\
\hline	 
$\varepsilon_{\rm HF}$ & $ -7.8646$ & $ -0.0000$ & $ -0.0000$ & $ -0.0000$ & $ -0.0000$ & $ -0.0000$ \\
$\varepsilon_U^{(2)}$ & $ -0.5155$ & $  0.0000$ & $  0.0000$ & $  0.0000$ & $  0.0000$ & $  0.0000$ \\
$\varepsilon_V^{(2)}$ & $  7.6330$ & $ -0.0050$ & $ -0.0134$ & $ -0.0351$ & $ -0.0905$ & $ -0.2273$ \\
\hline
\multicolumn{7}{c}{$k= 31\pi/249$}\\
\hline
         $M$ & \multicolumn{1}{c}{$\infty$}& \multicolumn{1}{c}{$   80676$} & \multicolumn{1}{c}{$   26892$} & \multicolumn{1}{c}{$    8964$} & \multicolumn{1}{c}{$    2988$} & \multicolumn{1}{c}{$     996$} \\
\hline	 
$\varepsilon_{\rm HF}$ & $ -6.8531$ & $ -0.0000$ & $ -0.0000$ & $ -0.0000$ & $ -0.0000$ & $ -0.0000$ \\
$\varepsilon_U^{(2)}$ & $ -0.8135$ & $  0.0000$ & $  0.0000$ & $  0.0000$ & $  0.0000$ & $  0.0000$ \\
$\varepsilon_V^{(2)}$ & $  7.3474$ & $ -0.0050$ & $ -0.0134$ & $ -0.0354$ & $ -0.0916$ & $ -0.2317$ \\
\hline
\multicolumn{7}{c}{$k= 62\pi/249$}\\
\hline
         $M$ & \multicolumn{1}{c}{$\infty$}& \multicolumn{1}{c}{$   80676$} & \multicolumn{1}{c}{$   26892$} & \multicolumn{1}{c}{$    8964$} & \multicolumn{1}{c}{$    2988$} & \multicolumn{1}{c}{$     996$} \\
\hline	 
$\varepsilon_{\rm HF}$ & $ -3.0542$ & $ -0.0000$ & $ -0.0000$ & $ -0.0000$ & $ -0.0003$ & $ -0.0003$ \\
$\varepsilon_U^{(2)}$ & $ -2.5380$ & $  0.0000$ & $  0.0000$ & $  0.0001$ & $  0.0013$ & $  0.0029$ \\
$\varepsilon_V^{(2)}$ & $  2.8929$ & $ -0.0077$ & $ -0.0235$ & $ -0.0694$ & $ -0.1725$ & $ -0.4138$ \\
\end{tabular}
\end{ruledtabular}
\end{table}

\begin{figure}
\includegraphics{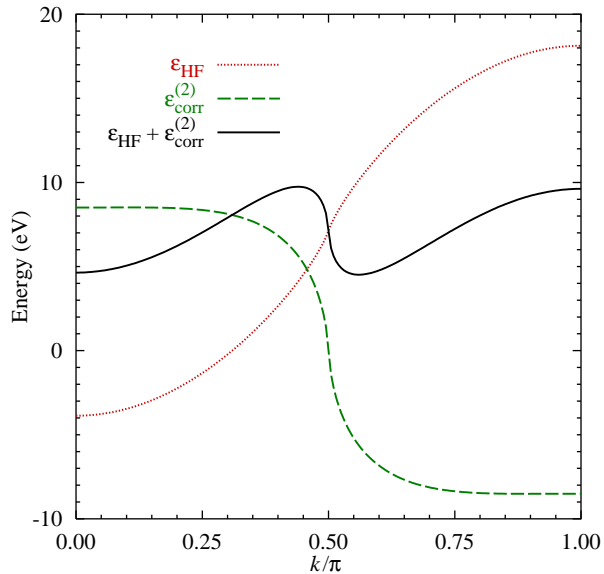}%
\caption{(Color online) \label{fig:PPP-Peps(k)pi/2}
PPP-P model, $k_{\rm F} = \pi/2$. Hartree-Fock orbital energies, $\varepsilon_{\rm HF}(k)$, corresponding second-order perturbation theory corrections,
 $\varepsilon_{\rm corr}^{(2)}(k)$, and their sum, shown as
functions of $k/\pi$.}
\end{figure}

\begin{figure}
\includegraphics{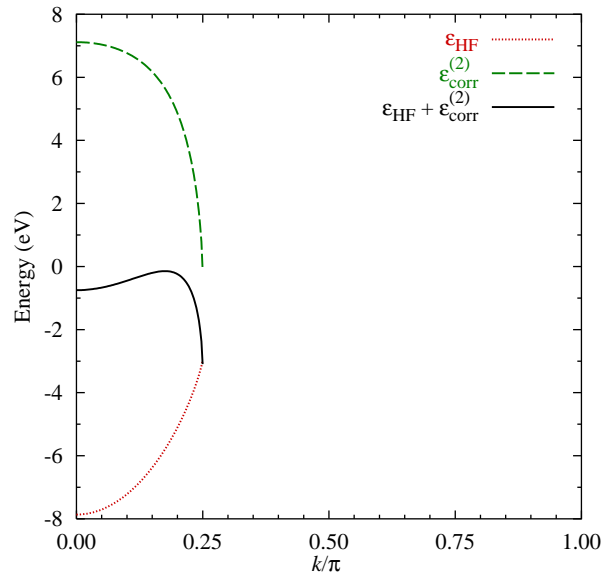}%
\caption{(Color online) \label{fig:PPP-Peps(k)pi/4}
PPP-P model, $k_{\rm F} = \pi/4$. Hartree-Fock  orbital energies, $\varepsilon_{\rm HF}(k)$, corresponding second-order perturbation theory corrections corrections, $\varepsilon_{\rm corr}^{(2)}(k)$, and their sum, shown as
functions of $k/\pi$.}
\end{figure}

In the case of a normal metal, one expects that the exact correlation corrections to the energy bands, $\varepsilon_{\rm corr}(k)$, should (i) reduce the width of the HF
band, $\varepsilon_{\rm HF}(k)$, and (ii) counterbalance its infinite slope at $k = \pm k_{\rm F}$. The second-order corrections, $\varepsilon_{\rm corr}^{(2)}(k)$, corresponding to the
PPP-P model of the 1D metal look as they fulfill (i) and (ii) at a qualitative level. However, the results in Figs.~\ref{fig:PPP-Peps(k)pi/2} and~\ref{fig:PPP-Peps(k)pi/4}
show that $\varepsilon_{\rm corr}^{(2)}(k)$, in fact, grossly overcorrect the deficiencies of the HF approximation. One may wonder, which theory level for $\varepsilon_{\rm corr}(k)$ is necessary to make conditions (i) and (ii) satisfactorily fulfilled. Aissing and Monkhorst~\cite{aissing:93} advocate using the Green's function approach of Ref.~\onlinecite{hedin:69}, within the random-phase approximation (RPA). Another promising candidate is the Fock-space coupled-cluster approach (see, e.g., Ref.~\onlinecite{mukherjee:89}, and references cited therein), employed by Emrich and Zabolitzky in their calculations of the single-particle energies corresponding to the
3D electron-gas model.~\cite{emrich:84} According to their analysis, the exact $\varepsilon_{\rm corr}(k)$ may be expressed as a sum of three terms,
$\varepsilon_{\rm corr}^{\rm (a)}(k)$,
 $\varepsilon_{\rm corr}^{\rm (b)}(k)$, and  $\varepsilon_{\rm corr}^{\rm (c)}(k)$
corresponding to diagrams depicted in Fig.~10 of Ref.~\onlinecite{emrich:84}. At the second-order level, the sum of $\varepsilon_{\rm corr}^{\rm (a)}(k)$
and $\varepsilon_{\rm corr}^{\rm (b)}(k)$ is equal to $\varepsilon_{\rm corr}^{(2)}(k)$ of our Eq.~(\ref{eq:eps2(k)}), while the lowest-order contribution to $\varepsilon_{\rm
corr}^{\rm (c)}(k)$ corresponds to the third-order  level. Emrich and Zabolitzky calculated the sum $\varepsilon_{\rm corr}^{\rm (a)}(k)+ \varepsilon_{\rm corr}^{\rm (b)}(k)$ within the
CC theory, and found that the derivative of this quantity, calculated at $k = k_{\rm F}$, exhibits a negative logarithmic singularity twice the strength of the positive
logarithmic singularity corresponding to the Hartree-Fock band energy, $\varepsilon_{\rm HF}(k)$. Only when they included term $\varepsilon_{\rm corr}^{\rm (c)}(k)$, the singularity
corresponding to $\varepsilon_{\rm HF}(k)$ was exactly canceled. Therefore, even at the coupled-cluster level of theory, with $\varepsilon(k)$ approximated by the sum $\varepsilon_{\rm HF}(k)
+ \varepsilon_{\rm corr}^{\rm (a)}(k) + \varepsilon_{\rm corr}^{\rm (b)}(k)$, the results for the 3D electron-gas model would look similarly to our
Fig.~\ref{fig:PPP-Peps(k)pi/2}. However, it seems that the problem cannot be fixed just by adding the third-order corrections (it is unlikely that a finite-order treatment is capable of the exact cancellation of the above-mentioned derivative singularity).

   The PPP model with the Pople potential (\ref{eq:gammaP(R)},\ref{eq:gamma0P}) takes into account the essential long-range electrostatic interactions in a 1D metal. Any other single-band model, including an {\em ab-initio\/} one, would provide only some short-range corrections to the PPP-P model, see Eqs.~(\ref{eq:gamma(R)}) and (\ref{eq:lambda(R)}). Therefore, we are confident that the results presented in this section give a {\em qualitatively correct\/} picture of the second-order correlation corrections to the electronic-energy bands of a 1D-metal.

\section{\label{sec:disc}%
Discussion}

  Let us recapitulate the most important findings of this paper:
\begin{itemize}
 \item Functions $F^{(2)}(q)$, corresponding to the second-order correlation energy per molecule (\ref{eq:E2corr}), and the second-order corrections to the energy bands, $\varepsilon_{\rm
corr}^{(2)}(k)$, calculated for the PPP model of 1D metal, display a striking similarity to the analogous functions calculated at the CC level of theory for the 3D
electron-gas model.~\cite{emrich:84}
 \item Functions $F^{(2)}(q)$ and $\varepsilon_{\rm corr}^{(2)}(k)$, calculated for the Hubbard-$0$ model of 1D metal are very different from those corresponding to the PPP
model.
\end{itemize}
We attribute the qualitative differences between the PPP and Hubbard-$0$ second-order results to the neglect of the long-range Coulomb interactions in the Hubbard-$0$ model. On the
other hand, the basic common element of the 3D electron-gas model and the PPP model of 1D metal is the absence of {\it a priori} screening of the Coulomb interactions.
Therefore, we treat the similarity of the results corresponding to these two models as a signature of the long-range effects. Within the range of electron
densities studied in Ref.~\onlinecite{emrich:84}, the 3D electron gas behaves as the paramagnetic Fermi fluid,~\cite{ceperley:80} and thus it may serve as a model of the
normal metal. We conclude that the second-order results of the present paper are consistent with the normal-metal picture for the PPP model of 1D metal.

   However, a different picture emerges when looking at the properties of the closed-shell ground state of an annulene described within the PPP model (which may be considered a special case of the PPP model of finite X$_M$ system):
\begin{itemize}
 \item The coupled-cluster (CC) methodology breaks down in the strongly correlated regime of the model (see Sec.~\ref{sec:intro}), due to the unusually strong coupling between the
connected-excitation operators of different ranks which contribute to the CC operator appearing in Eq.~(\ref{eq:exp(T)}).~\cite{podeszwa:02a} It seems that any variant of
the CC method employing a truncated CC operator will be ineffective for these systems. Such a behavior of the CC method is not consistent with the normal-metal character
of the annulene ground state (in the extended-system limit).
 \item The density-matrix renormalization group  calculations of various correlation functions for the ground state of the model~\cite{fano:99} indicate that it corresponds to the 1D analog of the Wigner
crystal, rather than to the normal-metal state, in agreement with the earlier predictions by Schulz.~\cite{schulz:93} The above-mentioned ineffectiveness of the CC method
may be just a consequence of the Wigner-crystal character of the ground state, and thus a failure of the Hartree-Fock wavefunction $\Phi_{\rm HF}$ as the reference function in
Eq.~(\ref{eq:exp(T)}).
\end{itemize}
One may thus conclude that there are strong indications against the normal-metal character of the ground state for the PPP model of 1D metal.

    The apparent contradiction between the conclusions of the above two paragraphs may be resolved by stating a hypothesis that the normal-metal state of our PPP model of
1D metal is different from the ground state, i.e., it corresponds to a certain excited state of the same symmetry (spin singlet, $k=0$). The eigenfunction of
Hamiltonian~(\ref{eq:PPPH(1)}) corresponding to this excited state, $\Psi_0^{'}$, should be more similar to $\Phi_{\rm HF}$ than is the ground-state eigenfunction,
$\Psi_0$. In such a case, one would find that the second-order corrections considered in this paper corresponded to the electronic-correlation effects associated with $\Psi_0^{'}$
(and the related ionized eigenstates), rather than to $\Psi_0$ (and the ionized eigenstates related to it). In our earlier studies~\cite{podeszwa:02d,podeszwa:03} we found some indirect evidence supporting the above hypothesis:
\begin{itemize}
 \item In Ref.~\onlinecite{podeszwa:03} we studied the PPP model of benzene ($M=6$), which, at the FCI level, has 18 eigenstates of the $\rm {}^1A_{1g}^{-}$ symmetry (the ${\rm D}_{6h}$ point group plus the alternancy symmetry). Our (unpublished) FCI calculations for the PPP-P model with $\beta^0 = -0.3$~eV (i.e., in the strongly correlated regime) revealed that the {\em fourth\/} eigenstate $\Psi_0^{'}$ corresponds to $\langle \Psi_0^{'} | \Phi_{\rm HF}\rangle = 0.62$, while for the ground state $\langle \Psi_0 | \Phi_{\rm HF}\rangle = 0.45$.
 \item When studying the multiple solutions of the CCD equations for the PPP-P model of [10]annulene~\cite{podeszwa:02d} ($M=10$), we found two CCD solutions: the ``standard'' one, describing the ground state in the weakly correlated regime, and the ``nonstandard'' one, which seemed to better reproduce the ground-state FCI energy in the strongly correlated regime (however, in this region, both solutions displayed little similarity to the ground-state FCI solution). Interestingly, when the CCD iterations used the second-order amplitudes~(\ref{eq:t2(1)}) as the starting point, they converged to the ``standard'' solution in the weakly correlated regime, while in the strongly correlated regime the ``nonstandard'' solution was obtained.  Unfortunately, a large dimension of the FCI-solution space has precluded the study of the similarity between the ``nonstandard'' solution and the excited FCI states.
\end{itemize}
Our ``hidden normal-metal state'' hypothesis is consistent with the properties of most quasi-1D conductors, which undergo an insulator-to-metal transition above $0$ K. In order to verify this hypothesis, a new strategy of solving the CC equations has to be developed: instead of looking for the solution corresponding to the lowest eigenstate of a given symmetry, one should focus on finding the solution most similar to the reference determinantal function. Adopting this strategy in the CC calculations for the 1D metal model may be effected by imposing certain restrictions on the $t(k_1,k_2,q)$-amplitudes of Eq.~(\ref{eq:tabij}); the CC calculations for the 3D electron gas\cite{freeman:77,bishop:78,bishop:82} may provide some guidance on the form of these restrictions. This route is now being investigated by us. 

  A truly 1D metallic system represents an extreme case of a highly anisotropic metal. In such a system, a dynamical screening of the electron-electron repulsion (the primary electronic-correlation effect in a 3D isotropic metal) is likely to be incomplete. That puts in doubt the suitability of the Hubbard model for the 1D metal. On the other hand, the Pariser-Parr-Pople model studied in the present paper captures the essential physics of the long-range Coulomb interactions in the 1D metal, without introducing an excessive computational overhead. Despite their different provenance, the PPP-P and the electron-gas models have much in common when applied to the description of the 1D and 3D metals, respectively. Further studies of the PPP-P model of 1D metal should help in clarifying the similarities and differences of the electronic-correlation effects in the 1D and 3D metallic systems.
\begin{acknowledgments}
The work was supported in part by the Committee for Scientific Research (KBN) through grant  7 T09A  019 20.
The authors are grateful to Robert Moszy\'{n}ski and Hendrik
Monkhorst for reading the manuscript and providing valuable comments.
\end{acknowledgments}

\bibliography{1D-2oec}

\begin{thebibliography}{109}
\expandafter\ifx\csname natexlab\endcsname\relax\def\natexlab#1{#1}\fi
\expandafter\ifx\csname bibnamefont\endcsname\relax
  \def\bibnamefont#1{#1}\fi
\expandafter\ifx\csname bibfnamefont\endcsname\relax
  \def\bibfnamefont#1{#1}\fi
\expandafter\ifx\csname citenamefont\endcsname\relax
  \def\citenamefont#1{#1}\fi
\expandafter\ifx\csname url\endcsname\relax
  \def\url#1{\texttt{#1}}\fi
\expandafter\ifx\csname urlprefix\endcsname\relax\def\urlprefix{URL }\fi
\providecommand{\bibinfo}[2]{#2}
\providecommand{\eprint}[2][]{\url{#2}}

\bibitem[{\citenamefont{Hedin and Lundquist}(1969)}]{hedin:69}
\bibinfo{author}{\bibfnamefont{L.}~\bibnamefont{Hedin}} \bibnamefont{and}
  \bibinfo{author}{\bibfnamefont{S.}~\bibnamefont{Lundquist}},
  \bibinfo{journal}{Solid State Phys.} \textbf{\bibinfo{volume}{23}},
  \bibinfo{pages}{1} (\bibinfo{year}{1969}).

\bibitem[{\citenamefont{Hubbard}(1963)}]{hubbard:63}
\bibinfo{author}{\bibfnamefont{J.}~\bibnamefont{Hubbard}},
  \bibinfo{journal}{Proc. R. Soc. London, Ser. A}
  \textbf{\bibinfo{volume}{276}}, \bibinfo{pages}{238} (\bibinfo{year}{1963}).

\bibitem[{\citenamefont{Lieb and Wu}(1968)}]{lieb:68}
\bibinfo{author}{\bibfnamefont{E.~H.} \bibnamefont{Lieb}} \bibnamefont{and}
  \bibinfo{author}{\bibfnamefont{F.~Y.} \bibnamefont{Wu}},
  \bibinfo{journal}{Phys. Rev. Lett.} \textbf{\bibinfo{volume}{20}},
  \bibinfo{pages}{1445} (\bibinfo{year}{1968}).

\bibitem[{\citenamefont{Go{\~{n}}i et~al.}(1991)\citenamefont{Go{\~{n}}i,
  Pinczuk, Weiner, Calleja, Dennis, Pfeiffer, and West}}]{goni:91}
\bibinfo{author}{\bibfnamefont{A.~R.} \bibnamefont{Go{\~{n}}i}},
  \bibinfo{author}{\bibfnamefont{A.}~\bibnamefont{Pinczuk}},
  \bibinfo{author}{\bibfnamefont{J.~S.} \bibnamefont{Weiner}},
  \bibinfo{author}{\bibfnamefont{J.~M.} \bibnamefont{Calleja}},
  \bibinfo{author}{\bibfnamefont{B.~S.} \bibnamefont{Dennis}},
  \bibinfo{author}{\bibfnamefont{L.~N.} \bibnamefont{Pfeiffer}},
  \bibnamefont{and} \bibinfo{author}{\bibfnamefont{K.~W.} \bibnamefont{West}},
  \bibinfo{journal}{Phys. Rev. Lett.} \textbf{\bibinfo{volume}{67}},
  \bibinfo{pages}{3298} (\bibinfo{year}{1991}).

\bibitem[{\citenamefont{Schulz}(1993)}]{schulz:93}
\bibinfo{author}{\bibfnamefont{H.~J.} \bibnamefont{Schulz}},
  \bibinfo{journal}{Phys. Rev. Lett.} \textbf{\bibinfo{volume}{71}},
  \bibinfo{pages}{1864} (\bibinfo{year}{1993}).

\bibitem[{\citenamefont{Poilblanc et~al.}(1997)\citenamefont{Poilblanc, Yunoki,
  Maekawa, and Dagotto}}]{poilblanc:97}
\bibinfo{author}{\bibfnamefont{D.}~\bibnamefont{Poilblanc}},
  \bibinfo{author}{\bibfnamefont{S.}~\bibnamefont{Yunoki}},
  \bibinfo{author}{\bibfnamefont{S.}~\bibnamefont{Maekawa}}, \bibnamefont{and}
  \bibinfo{author}{\bibfnamefont{E.}~\bibnamefont{Dagotto}},
  \bibinfo{journal}{Phys. Rev. B} \textbf{\bibinfo{volume}{56}},
  \bibinfo{pages}{R1645} (\bibinfo{year}{1997}).

\bibitem[{\citenamefont{Hatsugai}(1997)}]{hatsugai:97}
\bibinfo{author}{\bibfnamefont{Y.}~\bibnamefont{Hatsugai}},
  \bibinfo{journal}{Phys. Rev. B} \textbf{\bibinfo{volume}{56}},
  \bibinfo{pages}{12183} (\bibinfo{year}{1997}).

\bibitem[{\citenamefont{Fano et~al.}(1999)\citenamefont{Fano, Ortolani, Parola,
  and Ziosi}}]{fano:99}
\bibinfo{author}{\bibfnamefont{G.}~\bibnamefont{Fano}},
  \bibinfo{author}{\bibfnamefont{F.}~\bibnamefont{Ortolani}},
  \bibinfo{author}{\bibfnamefont{A.}~\bibnamefont{Parola}}, \bibnamefont{and}
  \bibinfo{author}{\bibfnamefont{L.}~\bibnamefont{Ziosi}},
  \bibinfo{journal}{Phys. Rev. B} \textbf{\bibinfo{volume}{60}},
  \bibinfo{pages}{15654} (\bibinfo{year}{1999}).

\bibitem[{\citenamefont{Capponi et~al.}(2000)\citenamefont{Capponi, Poilblanc,
  and Giamarchi}}]{capponi:00}
\bibinfo{author}{\bibfnamefont{S.}~\bibnamefont{Capponi}},
  \bibinfo{author}{\bibfnamefont{D.}~\bibnamefont{Poilblanc}},
  \bibnamefont{and}
  \bibinfo{author}{\bibfnamefont{T.}~\bibnamefont{Giamarchi}},
  \bibinfo{journal}{Phys. Rev. B} \textbf{\bibinfo{volume}{61}},
  \bibinfo{pages}{13410} (\bibinfo{year}{2000}).

\bibitem[{\citenamefont{Wang et~al.}(2001)\citenamefont{Wang, Millis, and
  Das~Sarma}}]{wang:01}
\bibinfo{author}{\bibfnamefont{D.~W.} \bibnamefont{Wang}},
  \bibinfo{author}{\bibfnamefont{A.~J.} \bibnamefont{Millis}},
  \bibnamefont{and}
  \bibinfo{author}{\bibfnamefont{S.}~\bibnamefont{Das~Sarma}},
  \bibinfo{journal}{Phys. Rev. B} \textbf{\bibinfo{volume}{64}},
  \bibinfo{pages}{193307} (\bibinfo{year}{2001}).

\bibitem[{\citenamefont{Lee}(2002)}]{lee:02}
\bibinfo{author}{\bibfnamefont{H.~C.} \bibnamefont{Lee}},
  \bibinfo{journal}{Phys. Rev. B} \textbf{\bibinfo{volume}{66}},
  \bibinfo{pages}{052202} (\bibinfo{year}{2002}).

\bibitem[{\citenamefont{Valenzuela et~al.}(2003)\citenamefont{Valenzuela,
  Fratini, and Baeriswyl}}]{valenzuela:03}
\bibinfo{author}{\bibfnamefont{B.}~\bibnamefont{Valenzuela}},
  \bibinfo{author}{\bibfnamefont{S.}~\bibnamefont{Fratini}}, \bibnamefont{and}
  \bibinfo{author}{\bibfnamefont{D.}~\bibnamefont{Baeriswyl}},
  \bibinfo{journal}{Phys. Rev. B} \textbf{\bibinfo{volume}{68}},
  \bibinfo{pages}{045112} (\bibinfo{year}{2003}).

\bibitem[{\citenamefont{M{\o}ller and Plesset}(1934)}]{moller:34}
\bibinfo{author}{\bibfnamefont{C.}~\bibnamefont{M{\o}ller}} \bibnamefont{and}
  \bibinfo{author}{\bibfnamefont{M.~S.} \bibnamefont{Plesset}},
  \bibinfo{journal}{Phys. Rev.} \textbf{\bibinfo{volume}{46}},
  \bibinfo{pages}{618} (\bibinfo{year}{1934}).

\bibitem[{\citenamefont{Brueckner}(1955)}]{brueckner:55}
\bibinfo{author}{\bibfnamefont{K.~A.} \bibnamefont{Brueckner}},
  \bibinfo{journal}{Phys. Rev.} \textbf{\bibinfo{volume}{100}},
  \bibinfo{pages}{36} (\bibinfo{year}{1955}).

\bibitem[{\citenamefont{Goldstone}(1957)}]{goldstone:57}
\bibinfo{author}{\bibfnamefont{J.}~\bibnamefont{Goldstone}},
  \bibinfo{journal}{Proc. R. Soc. London, Ser A}
  \textbf{\bibinfo{volume}{239}}, \bibinfo{pages}{267} (\bibinfo{year}{1957}).

\bibitem[{\citenamefont{Bartlett}(1981)}]{bartlett:81}
\bibinfo{author}{\bibfnamefont{R.~J.} \bibnamefont{Bartlett}},
  \bibinfo{journal}{Annu. Rev. Phys. Chem.} \textbf{\bibinfo{volume}{32}},
  \bibinfo{pages}{359} (\bibinfo{year}{1981}).

\bibitem[{\citenamefont{Bryce and Murphy}(1984)}]{bryce:84}
\bibinfo{author}{\bibfnamefont{M.~R.} \bibnamefont{Bryce}} \bibnamefont{and}
  \bibinfo{author}{\bibfnamefont{L.~C.} \bibnamefont{Murphy}},
  \bibinfo{journal}{Nature} \textbf{\bibinfo{volume}{309}},
  \bibinfo{pages}{119} (\bibinfo{year}{1984}).

\bibitem[{\citenamefont{Pariser and Parr}(1953{\natexlab{a}})}]{pariser:53a}
\bibinfo{author}{\bibfnamefont{R.}~\bibnamefont{Pariser}} \bibnamefont{and}
  \bibinfo{author}{\bibfnamefont{R.}~\bibnamefont{Parr}}, \bibinfo{journal}{J.
  Chem. Phys.} \textbf{\bibinfo{volume}{21}}, \bibinfo{pages}{466}
  (\bibinfo{year}{1953}{\natexlab{a}}).

\bibitem[{\citenamefont{Pariser and Parr}(1953{\natexlab{b}})}]{pariser:53b}
\bibinfo{author}{\bibfnamefont{R.}~\bibnamefont{Pariser}} \bibnamefont{and}
  \bibinfo{author}{\bibfnamefont{R.}~\bibnamefont{Parr}}, \bibinfo{journal}{J.
  Chem. Phys.} \textbf{\bibinfo{volume}{21}}, \bibinfo{pages}{767}
  (\bibinfo{year}{1953}{\natexlab{b}}).

\bibitem[{\citenamefont{Pople}(1953)}]{pople:53}
\bibinfo{author}{\bibfnamefont{J.~A.} \bibnamefont{Pople}},
  \bibinfo{journal}{Trans. Faraday Soc.} \textbf{\bibinfo{volume}{49}},
  \bibinfo{pages}{1375} (\bibinfo{year}{1953}).

\bibitem[{\citenamefont{Fano et~al.}(1998)\citenamefont{Fano, Ortolani, and
  Ziosi}}]{fano:98}
\bibinfo{author}{\bibfnamefont{G.}~\bibnamefont{Fano}},
  \bibinfo{author}{\bibfnamefont{F.}~\bibnamefont{Ortolani}}, \bibnamefont{and}
  \bibinfo{author}{\bibfnamefont{L.}~\bibnamefont{Ziosi}}, \bibinfo{journal}{J.
  Chem. Phys.} \textbf{\bibinfo{volume}{108}}, \bibinfo{pages}{9246}
  (\bibinfo{year}{1998}).

\bibitem[{\citenamefont{Piela and Delhalle}(1978)}]{piela:78}
\bibinfo{author}{\bibfnamefont{L.}~\bibnamefont{Piela}} \bibnamefont{and}
  \bibinfo{author}{\bibfnamefont{J.}~\bibnamefont{Delhalle}},
  \bibinfo{journal}{Int. J. Quantum Chem.} \textbf{\bibinfo{volume}{13}},
  \bibinfo{pages}{605} (\bibinfo{year}{1978}).

\bibitem[{\citenamefont{Delhalle et~al.}(1980)\citenamefont{Delhalle, Piela,
  Br{\'e}das, and Andr{\'e}}}]{delhalle:80}
\bibinfo{author}{\bibfnamefont{J.}~\bibnamefont{Delhalle}},
  \bibinfo{author}{\bibfnamefont{L.}~\bibnamefont{Piela}},
  \bibinfo{author}{\bibfnamefont{J.-L.} \bibnamefont{Br{\'e}das}},
  \bibnamefont{and} \bibinfo{author}{\bibfnamefont{J.-M.}
  \bibnamefont{Andr{\'e}}}, \bibinfo{journal}{Phys. Rev. B}
  \textbf{\bibinfo{volume}{22}}, \bibinfo{pages}{6254} (\bibinfo{year}{1980}).

\bibitem[{\citenamefont{Stolarczyk et~al.}(1988)\citenamefont{Stolarczyk,
  Jeziorska, and Monkhorst}}]{stolarczyk:88}
\bibinfo{author}{\bibfnamefont{L.~Z.} \bibnamefont{Stolarczyk}},
  \bibinfo{author}{\bibfnamefont{M.}~\bibnamefont{Jeziorska}},
  \bibnamefont{and} \bibinfo{author}{\bibfnamefont{H.~J.}
  \bibnamefont{Monkhorst}}, \bibinfo{journal}{Phys. Rev. B}
  \textbf{\bibinfo{volume}{37}}, \bibinfo{pages}{10646} (\bibinfo{year}{1988}).

\bibitem[{\citenamefont{Jeziorska et~al.}(1990)\citenamefont{Jeziorska,
  Stolarczyk, Paldus, and Monkhorst}}]{jeziorska:90}
\bibinfo{author}{\bibfnamefont{M.}~\bibnamefont{Jeziorska}},
  \bibinfo{author}{\bibfnamefont{L.~Z.} \bibnamefont{Stolarczyk}},
  \bibinfo{author}{\bibfnamefont{J.}~\bibnamefont{Paldus}}, \bibnamefont{and}
  \bibinfo{author}{\bibfnamefont{H.~J.} \bibnamefont{Monkhorst}},
  \bibinfo{journal}{Phys. Rev. B} \textbf{\bibinfo{volume}{41}},
  \bibinfo{pages}{12473} (\bibinfo{year}{1990}).

\bibitem[{\citenamefont{Monkhorst}(1979)}]{monkhorst:79}
\bibinfo{author}{\bibfnamefont{H.~J.} \bibnamefont{Monkhorst}},
  \bibinfo{journal}{Phys. Rev. B} \textbf{\bibinfo{volume}{20}},
  \bibinfo{pages}{1504} (\bibinfo{year}{1979}).

\bibitem[{\citenamefont{Coester}(1958)}]{coester:58}
\bibinfo{author}{\bibfnamefont{F.}~\bibnamefont{Coester}},
  \bibinfo{journal}{Nucl. Phys.} \textbf{\bibinfo{volume}{7}},
  \bibinfo{pages}{421} (\bibinfo{year}{1958}).

\bibitem[{\citenamefont{{{\v C}{\' \i}{\v{z}ek}}}(1966)}]{cizek:66}
\bibinfo{author}{\bibfnamefont{J.}~\bibnamefont{{{\v C}{\' \i}{\v{z}ek}}}},
  \bibinfo{journal}{J. Chem. Phys.} \textbf{\bibinfo{volume}{45}},
  \bibinfo{pages}{4256} (\bibinfo{year}{1966}).

\bibitem[{\citenamefont{Suhai and Ladik}(1982)}]{suhai:82}
\bibinfo{author}{\bibfnamefont{S.}~\bibnamefont{Suhai}} \bibnamefont{and}
  \bibinfo{author}{\bibfnamefont{J.}~\bibnamefont{Ladik}}, \bibinfo{journal}{J.
  Phys. C} \textbf{\bibinfo{volume}{15}}, \bibinfo{pages}{4327}
  (\bibinfo{year}{1982}).

\bibitem[{\citenamefont{Suhai}(1983{\natexlab{a}})}]{suhai:83a}
\bibinfo{author}{\bibfnamefont{S.}~\bibnamefont{Suhai}},
  \bibinfo{journal}{Phys. Rev. B} \textbf{\bibinfo{volume}{27}},
  \bibinfo{pages}{3506} (\bibinfo{year}{1983}{\natexlab{a}}).

\bibitem[{\citenamefont{Suhai}(1983{\natexlab{b}})}]{suhai:83b}
\bibinfo{author}{\bibfnamefont{S.}~\bibnamefont{Suhai}},
  \bibinfo{journal}{Chem. Phys. Lett.} \textbf{\bibinfo{volume}{96}},
  \bibinfo{pages}{619} (\bibinfo{year}{1983}{\natexlab{b}}).

\bibitem[{\citenamefont{Suhai}(1983{\natexlab{c}})}]{suhai:83c}
\bibinfo{author}{\bibfnamefont{S.}~\bibnamefont{Suhai}}, \bibinfo{journal}{Int.
  J. Quantum Chem.} \textbf{\bibinfo{volume}{23}}, \bibinfo{pages}{1239}
  (\bibinfo{year}{1983}{\natexlab{c}}).

\bibitem[{\citenamefont{Suhai}(1992)}]{suhai:92}
\bibinfo{author}{\bibfnamefont{S.}~\bibnamefont{Suhai}}, \bibinfo{journal}{Int.
  J. Quantum Chem.} \textbf{\bibinfo{volume}{42}}, \bibinfo{pages}{193}
  (\bibinfo{year}{1992}).

\bibitem[{\citenamefont{Suhai}(1993)}]{suhai:93}
\bibinfo{author}{\bibfnamefont{S.}~\bibnamefont{Suhai}}, \bibinfo{journal}{Int.
  J. Quantum Chem. (Quantum Chem. Symposium)} \textbf{\bibinfo{volume}{27}},
  \bibinfo{pages}{131} (\bibinfo{year}{1993}).

\bibitem[{\citenamefont{Suhai}(1994{\natexlab{a}})}]{suhai:94a}
\bibinfo{author}{\bibfnamefont{S.}~\bibnamefont{Suhai}},
  \bibinfo{journal}{Phys. Rev. B} \textbf{\bibinfo{volume}{50}},
  \bibinfo{pages}{14791} (\bibinfo{year}{1994}{\natexlab{a}}).

\bibitem[{\citenamefont{Suhai}(1994{\natexlab{b}})}]{suhai:94b}
\bibinfo{author}{\bibfnamefont{S.}~\bibnamefont{Suhai}}, \bibinfo{journal}{J.
  Chem. Phys.} \textbf{\bibinfo{volume}{101}}, \bibinfo{pages}{9766}
  (\bibinfo{year}{1994}{\natexlab{b}}).

\bibitem[{\citenamefont{Suhai}(1995{\natexlab{a}})}]{suhai:95a}
\bibinfo{author}{\bibfnamefont{S.}~\bibnamefont{Suhai}},
  \bibinfo{journal}{Phys. Rev. B} \textbf{\bibinfo{volume}{51}},
  \bibinfo{pages}{16553} (\bibinfo{year}{1995}{\natexlab{a}}).

\bibitem[{\citenamefont{Suhai}(1995{\natexlab{b}})}]{suhai:95b}
\bibinfo{author}{\bibfnamefont{S.}~\bibnamefont{Suhai}},
  \bibinfo{journal}{Phys. Rev. B} \textbf{\bibinfo{volume}{52}},
  \bibinfo{pages}{1674} (\bibinfo{year}{1995}{\natexlab{b}}).

\bibitem[{\citenamefont{Sun and Bartlett}(1996{\natexlab{a}})}]{sun:96a}
\bibinfo{author}{\bibfnamefont{J.-Q.} \bibnamefont{Sun}} \bibnamefont{and}
  \bibinfo{author}{\bibfnamefont{R.~J.} \bibnamefont{Bartlett}},
  \bibinfo{journal}{J. Chem. Phys.} \textbf{\bibinfo{volume}{104}},
  \bibinfo{pages}{8553} (\bibinfo{year}{1996}{\natexlab{a}}).

\bibitem[{\citenamefont{Sun and Bartlett}(1996{\natexlab{b}})}]{sun:96b}
\bibinfo{author}{\bibfnamefont{J.-Q.} \bibnamefont{Sun}} \bibnamefont{and}
  \bibinfo{author}{\bibfnamefont{R.~J.} \bibnamefont{Bartlett}},
  \bibinfo{journal}{Phys. Rev. Lett.} \textbf{\bibinfo{volume}{77}},
  \bibinfo{pages}{3669} (\bibinfo{year}{1996}{\natexlab{b}}).

\bibitem[{\citenamefont{Sun and Bartlett}(1997)}]{sun:97}
\bibinfo{author}{\bibfnamefont{J.-Q.} \bibnamefont{Sun}} \bibnamefont{and}
  \bibinfo{author}{\bibfnamefont{R.~J.} \bibnamefont{Bartlett}},
  \bibinfo{journal}{J. Chem. Phys.} \textbf{\bibinfo{volume}{107}},
  \bibinfo{pages}{5058} (\bibinfo{year}{1997}).

\bibitem[{\citenamefont{Sun and Bartlett}(1998)}]{sun:98}
\bibinfo{author}{\bibfnamefont{J.-Q.} \bibnamefont{Sun}} \bibnamefont{and}
  \bibinfo{author}{\bibfnamefont{R.~J.} \bibnamefont{Bartlett}},
  \bibinfo{journal}{Phys. Rev. Lett.} \textbf{\bibinfo{volume}{80}},
  \bibinfo{pages}{349} (\bibinfo{year}{1998}).

\bibitem[{\citenamefont{Sun and Bartlett}(1999)}]{sun:99}
\bibinfo{author}{\bibfnamefont{J.-Q.} \bibnamefont{Sun}} \bibnamefont{and}
  \bibinfo{author}{\bibfnamefont{R.~J.} \bibnamefont{Bartlett}},
  \bibinfo{journal}{Top. Curr. Chem.} \textbf{\bibinfo{volume}{203}},
  \bibinfo{pages}{121} (\bibinfo{year}{1999}).

\bibitem[{\citenamefont{Hirata et~al.}(2001)\citenamefont{Hirata, Grabowski,
  Tobita, and Bartlett}}]{hirata:01}
\bibinfo{author}{\bibfnamefont{S.}~\bibnamefont{Hirata}},
  \bibinfo{author}{\bibfnamefont{I.}~\bibnamefont{Grabowski}},
  \bibinfo{author}{\bibfnamefont{M.}~\bibnamefont{Tobita}}, \bibnamefont{and}
  \bibinfo{author}{\bibfnamefont{R.~J.} \bibnamefont{Bartlett}},
  \bibinfo{journal}{Chem. Phys. Lett.} \textbf{\bibinfo{volume}{345}},
  \bibinfo{pages}{475} (\bibinfo{year}{2001}).

\bibitem[{\citenamefont{F{\"o}rner}(1992)}]{forner:92}
\bibinfo{author}{\bibfnamefont{W.}~\bibnamefont{F{\"o}rner}},
  \bibinfo{journal}{Int. J. Quantum Chem.} \textbf{\bibinfo{volume}{43}},
  \bibinfo{pages}{221} (\bibinfo{year}{1992}).

\bibitem[{\citenamefont{Ye et~al.}(1993)\citenamefont{Ye, F{\"o}rner, and
  Ladik}}]{ye:93}
\bibinfo{author}{\bibfnamefont{Y.-J.} \bibnamefont{Ye}},
  \bibinfo{author}{\bibfnamefont{W.}~\bibnamefont{F{\"o}rner}},
  \bibnamefont{and} \bibinfo{author}{\bibfnamefont{J.}~\bibnamefont{Ladik}},
  \bibinfo{journal}{Chem. Phys.} \textbf{\bibinfo{volume}{178}},
  \bibinfo{pages}{1} (\bibinfo{year}{1993}).

\bibitem[{\citenamefont{Knab et~al.}(1996)\citenamefont{Knab, F{\"o}rner, {{\v
  C}{\'\i}{\v{z}ek}}, and Ladik}}]{knab:96}
\bibinfo{author}{\bibfnamefont{R.}~\bibnamefont{Knab}},
  \bibinfo{author}{\bibfnamefont{W.}~\bibnamefont{F{\"o}rner}},
  \bibinfo{author}{\bibfnamefont{J.}~\bibnamefont{{{\v C}{\'\i}{\v{z}ek}}}},
  \bibnamefont{and} \bibinfo{author}{\bibfnamefont{J.}~\bibnamefont{Ladik}},
  \bibinfo{journal}{J. Mol. Struct. (Theochem)} \textbf{\bibinfo{volume}{366}},
  \bibinfo{pages}{11} (\bibinfo{year}{1996}).

\bibitem[{\citenamefont{Knab et~al.}(1997)\citenamefont{Knab, F{\"o}rner, and
  Ladik}}]{knab:97}
\bibinfo{author}{\bibfnamefont{R.}~\bibnamefont{Knab}},
  \bibinfo{author}{\bibfnamefont{W.}~\bibnamefont{F{\"o}rner}},
  \bibnamefont{and} \bibinfo{author}{\bibfnamefont{J.}~\bibnamefont{Ladik}},
  \bibinfo{journal}{J. Phys. Condens. Matter} \textbf{\bibinfo{volume}{9}},
  \bibinfo{pages}{2043} (\bibinfo{year}{1997}).

\bibitem[{\citenamefont{F{\"o}rner et~al.}(1997)\citenamefont{F{\"o}rner, Knab,
  {{\v C}{\'\i}{\v{z}ek}}, and Ladik}}]{forner:97}
\bibinfo{author}{\bibfnamefont{W.}~\bibnamefont{F{\"o}rner}},
  \bibinfo{author}{\bibfnamefont{R.}~\bibnamefont{Knab}},
  \bibinfo{author}{\bibfnamefont{J.}~\bibnamefont{{{\v C}{\'\i}{\v{z}ek}}}},
  \bibnamefont{and} \bibinfo{author}{\bibfnamefont{J.}~\bibnamefont{Ladik}},
  \bibinfo{journal}{J. Chem. Phys.} \textbf{\bibinfo{volume}{106}},
  \bibinfo{pages}{10248} (\bibinfo{year}{1997}).

\bibitem[{\citenamefont{Gell-Mann and Brueckner}(1957)}]{gell-mann:57a}
\bibinfo{author}{\bibfnamefont{M.}~\bibnamefont{Gell-Mann}} \bibnamefont{and}
  \bibinfo{author}{\bibfnamefont{K.~A.} \bibnamefont{Brueckner}},
  \bibinfo{journal}{Phys. Rev.} \textbf{\bibinfo{volume}{106}},
  \bibinfo{pages}{364} (\bibinfo{year}{1957}).

\bibitem[{\citenamefont{Gell-Mann}(1957)}]{gell-mann:57b}
\bibinfo{author}{\bibfnamefont{M.}~\bibnamefont{Gell-Mann}},
  \bibinfo{journal}{Phys. Rev.} \textbf{\bibinfo{volume}{106}},
  \bibinfo{pages}{369} (\bibinfo{year}{1957}).

\bibitem[{\citenamefont{Handler}(1988)}]{handler:88}
\bibinfo{author}{\bibfnamefont{G.~S.} \bibnamefont{Handler}},
  \bibinfo{journal}{Int. J. Quantum Chem.} \textbf{\bibinfo{volume}{33}},
  \bibinfo{pages}{173} (\bibinfo{year}{1988}).

\bibitem[{\citenamefont{Freeman}(1977)}]{freeman:77}
\bibinfo{author}{\bibfnamefont{D.~L.} \bibnamefont{Freeman}},
  \bibinfo{journal}{Phys. Rev. B} \textbf{\bibinfo{volume}{15}},
  \bibinfo{pages}{5512} (\bibinfo{year}{1977}).

\bibitem[{\citenamefont{Bishop and L{\"u}hrmann}(1978)}]{bishop:78}
\bibinfo{author}{\bibfnamefont{R.~F.} \bibnamefont{Bishop}} \bibnamefont{and}
  \bibinfo{author}{\bibfnamefont{K.~H.} \bibnamefont{L{\"u}hrmann}},
  \bibinfo{journal}{Phys. Rev. B} \textbf{\bibinfo{volume}{17}},
  \bibinfo{pages}{3757} (\bibinfo{year}{1978}).

\bibitem[{\citenamefont{Bishop and L{\"u}hrmann}(1982)}]{bishop:82}
\bibinfo{author}{\bibfnamefont{R.~F.} \bibnamefont{Bishop}} \bibnamefont{and}
  \bibinfo{author}{\bibfnamefont{K.~H.} \bibnamefont{L{\"u}hrmann}},
  \bibinfo{journal}{Phys. Rev. B} \textbf{\bibinfo{volume}{26}},
  \bibinfo{pages}{5523} (\bibinfo{year}{1982}).

\bibitem[{\citenamefont{Emrich and Zabolitzky}(1984)}]{emrich:84}
\bibinfo{author}{\bibfnamefont{K.}~\bibnamefont{Emrich}} \bibnamefont{and}
  \bibinfo{author}{\bibfnamefont{J.~G.} \bibnamefont{Zabolitzky}},
  \bibinfo{journal}{Phys. Rev. B} \textbf{\bibinfo{volume}{30}},
  \bibinfo{pages}{2049} (\bibinfo{year}{1984}).

\bibitem[{\citenamefont{Ceperley and Alder}(1980)}]{ceperley:80}
\bibinfo{author}{\bibfnamefont{D.~M.} \bibnamefont{Ceperley}} \bibnamefont{and}
  \bibinfo{author}{\bibfnamefont{B.~J.} \bibnamefont{Alder}},
  \bibinfo{journal}{Phys. Rev. Lett.} \textbf{\bibinfo{volume}{45}},
  \bibinfo{pages}{566} (\bibinfo{year}{1980}).

\bibitem[{\citenamefont{Paldus and Boyle}(1982)}]{paldus:82}
\bibinfo{author}{\bibfnamefont{J.}~\bibnamefont{Paldus}} \bibnamefont{and}
  \bibinfo{author}{\bibfnamefont{M.~J.} \bibnamefont{Boyle}},
  \bibinfo{journal}{Int. J. Quantum Chem.} \textbf{\bibinfo{volume}{22}},
  \bibinfo{pages}{1281} (\bibinfo{year}{1982}).

\bibitem[{\citenamefont{Takahashi et~al.}(1983)\citenamefont{Takahashi, Paldus,
  and {{\v C}{\' \i}{\v{z}ek}}}}]{takahashi:83}
\bibinfo{author}{\bibfnamefont{M.}~\bibnamefont{Takahashi}},
  \bibinfo{author}{\bibfnamefont{J.}~\bibnamefont{Paldus}}, \bibnamefont{and}
  \bibinfo{author}{\bibfnamefont{J.}~\bibnamefont{{{\v C}{\' \i}{\v{z}ek}}}},
  \bibinfo{journal}{Int. J. Quantum Chem.} \textbf{\bibinfo{volume}{24}},
  \bibinfo{pages}{707} (\bibinfo{year}{1983}).

\bibitem[{\citenamefont{Paldus et~al.}(1984{\natexlab{a}})\citenamefont{Paldus,
  Takahashi, and Cho}}]{paldus:84a}
\bibinfo{author}{\bibfnamefont{J.}~\bibnamefont{Paldus}},
  \bibinfo{author}{\bibfnamefont{M.}~\bibnamefont{Takahashi}},
  \bibnamefont{and} \bibinfo{author}{\bibfnamefont{R.~W.~H.}
  \bibnamefont{Cho}}, \bibinfo{journal}{Phys. Rev. B}
  \textbf{\bibinfo{volume}{30}}, \bibinfo{pages}{4267}
  (\bibinfo{year}{1984}{\natexlab{a}}).

\bibitem[{\citenamefont{Paldus et~al.}(1984{\natexlab{b}})\citenamefont{Paldus,
  {{\v C}{\' \i}{\v{z}ek}}, and Takahashi}}]{paldus:84b}
\bibinfo{author}{\bibfnamefont{J.}~\bibnamefont{Paldus}},
  \bibinfo{author}{\bibfnamefont{J.}~\bibnamefont{{{\v C}{\' \i}{\v{z}ek}}}},
  \bibnamefont{and}
  \bibinfo{author}{\bibfnamefont{M.}~\bibnamefont{Takahashi}},
  \bibinfo{journal}{Phys. Rev. A} \textbf{\bibinfo{volume}{30}},
  \bibinfo{pages}{2193} (\bibinfo{year}{1984}{\natexlab{b}}).

\bibitem[{\citenamefont{Paldus et~al.}(1984{\natexlab{c}})\citenamefont{Paldus,
  Takahashi, and Cho}}]{paldus:84c}
\bibinfo{author}{\bibfnamefont{J.}~\bibnamefont{Paldus}},
  \bibinfo{author}{\bibfnamefont{M.}~\bibnamefont{Takahashi}},
  \bibnamefont{and} \bibinfo{author}{\bibfnamefont{R.~W.~H.}
  \bibnamefont{Cho}}, \bibinfo{journal}{Int. J. Quantum Chem. (Quantum Chem.
  Symposium)} \textbf{\bibinfo{volume}{18}}, \bibinfo{pages}{237}
  (\bibinfo{year}{1984}{\natexlab{c}}).

\bibitem[{\citenamefont{Takahashi and Paldus}(1985)}]{takahashi:85}
\bibinfo{author}{\bibfnamefont{M.}~\bibnamefont{Takahashi}} \bibnamefont{and}
  \bibinfo{author}{\bibfnamefont{J.}~\bibnamefont{Paldus}},
  \bibinfo{journal}{Phys. Rev. B} \textbf{\bibinfo{volume}{31}},
  \bibinfo{pages}{5121} (\bibinfo{year}{1985}).

\bibitem[{\citenamefont{Piecuch and Paldus}(1990)}]{piecuch:90a}
\bibinfo{author}{\bibfnamefont{P.}~\bibnamefont{Piecuch}} \bibnamefont{and}
  \bibinfo{author}{\bibfnamefont{J.}~\bibnamefont{Paldus}},
  \bibinfo{journal}{Theor. Chim. Acta} \textbf{\bibinfo{volume}{78}},
  \bibinfo{pages}{65} (\bibinfo{year}{1990}).

\bibitem[{\citenamefont{Piecuch
  et~al.}(1990{\natexlab{a}})\citenamefont{Piecuch, Zarrabian, Paldus, and {{\v
  C}{\' \i}{\v{z}ek}}}}]{piecuch:90b}
\bibinfo{author}{\bibfnamefont{P.}~\bibnamefont{Piecuch}},
  \bibinfo{author}{\bibfnamefont{S.}~\bibnamefont{Zarrabian}},
  \bibinfo{author}{\bibfnamefont{J.}~\bibnamefont{Paldus}}, \bibnamefont{and}
  \bibinfo{author}{\bibfnamefont{J.}~\bibnamefont{{{\v C}{\' \i}{\v{z}ek}}}},
  \bibinfo{journal}{Phys. Rev. B} \textbf{\bibinfo{volume}{42}},
  \bibinfo{pages}{3351} (\bibinfo{year}{1990}{\natexlab{a}}).

\bibitem[{\citenamefont{Piecuch
  et~al.}(1990{\natexlab{b}})\citenamefont{Piecuch, Zarrabian, Paldus, and {{\v
  C}{\' \i}{\v{z}ek}}}}]{piecuch:90c}
\bibinfo{author}{\bibfnamefont{P.}~\bibnamefont{Piecuch}},
  \bibinfo{author}{\bibfnamefont{S.}~\bibnamefont{Zarrabian}},
  \bibinfo{author}{\bibfnamefont{J.}~\bibnamefont{Paldus}}, \bibnamefont{and}
  \bibinfo{author}{\bibfnamefont{J.}~\bibnamefont{{{\v C}{\' \i}{\v{z}ek}}}},
  \bibinfo{journal}{Phys. Rev. A} \textbf{\bibinfo{volume}{42}},
  \bibinfo{pages}{5155} (\bibinfo{year}{1990}{\natexlab{b}}).

\bibitem[{\citenamefont{Piecuch and Paldus}(1991)}]{piecuch:91}
\bibinfo{author}{\bibfnamefont{P.}~\bibnamefont{Piecuch}} \bibnamefont{and}
  \bibinfo{author}{\bibfnamefont{J.}~\bibnamefont{Paldus}},
  \bibinfo{journal}{Int. J. Quantum Chem. (Quantum Chem. Symposium)}
  \textbf{\bibinfo{volume}{25}}, \bibinfo{pages}{9} (\bibinfo{year}{1991}).

\bibitem[{\citenamefont{Paldus and Piecuch}(1992)}]{paldus:92}
\bibinfo{author}{\bibfnamefont{J.}~\bibnamefont{Paldus}} \bibnamefont{and}
  \bibinfo{author}{\bibfnamefont{P.}~\bibnamefont{Piecuch}},
  \bibinfo{journal}{Int. J. Quantum Chem.} \textbf{\bibinfo{volume}{42}},
  \bibinfo{pages}{135} (\bibinfo{year}{1992}).

\bibitem[{\citenamefont{Piecuch et~al.}(1992)\citenamefont{Piecuch, {{\v C}{\'
  \i}{\v{z}ek}}, and Paldus}}]{piecuch:92}
\bibinfo{author}{\bibfnamefont{P.}~\bibnamefont{Piecuch}},
  \bibinfo{author}{\bibfnamefont{J.}~\bibnamefont{{{\v C}{\' \i}{\v{z}ek}}}},
  \bibnamefont{and} \bibinfo{author}{\bibfnamefont{J.}~\bibnamefont{Paldus}},
  \bibinfo{journal}{Int. J. Quantum Chem.} \textbf{\bibinfo{volume}{42}},
  \bibinfo{pages}{165} (\bibinfo{year}{1992}).

\bibitem[{\citenamefont{Podeszwa et~al.}(2002)\citenamefont{Podeszwa,
  Kucharski, and Stolarczyk}}]{podeszwa:02a}
\bibinfo{author}{\bibfnamefont{R.}~\bibnamefont{Podeszwa}},
  \bibinfo{author}{\bibfnamefont{S.~A.} \bibnamefont{Kucharski}},
  \bibnamefont{and} \bibinfo{author}{\bibfnamefont{L.~Z.}
  \bibnamefont{Stolarczyk}}, \bibinfo{journal}{J. Chem. Phys.}
  \textbf{\bibinfo{volume}{116}}, \bibinfo{pages}{480} (\bibinfo{year}{2002}).

\bibitem[{\citenamefont{Podeszwa et~al.}(2003)\citenamefont{Podeszwa,
  Stolarczyk, Jankowski, and Rubiniec}}]{podeszwa:03}
\bibinfo{author}{\bibfnamefont{R.}~\bibnamefont{Podeszwa}},
  \bibinfo{author}{\bibfnamefont{L.~Z.} \bibnamefont{Stolarczyk}},
  \bibinfo{author}{\bibfnamefont{K.}~\bibnamefont{Jankowski}},
  \bibnamefont{and} \bibinfo{author}{\bibfnamefont{K.}~\bibnamefont{Rubiniec}},
  \bibinfo{journal}{Theor. Chem. Acc.} \textbf{\bibinfo{volume}{109}},
  \bibinfo{pages}{309} (\bibinfo{year}{2003}).

\bibitem[{\citenamefont{Podeszwa}(2002)}]{podeszwa:02c}
\bibinfo{author}{\bibfnamefont{R.}~\bibnamefont{Podeszwa}},
  \bibinfo{journal}{Chem. Phys. Lett.} \textbf{\bibinfo{volume}{365}},
  \bibinfo{pages}{211} (\bibinfo{year}{2002}).

\bibitem[{\citenamefont{Podeszwa and Stolarczyk}(2002)}]{podeszwa:02d}
\bibinfo{author}{\bibfnamefont{R.}~\bibnamefont{Podeszwa}} \bibnamefont{and}
  \bibinfo{author}{\bibfnamefont{L.~Z.} \bibnamefont{Stolarczyk}},
  \bibinfo{journal}{Chem. Phys. Lett.} \textbf{\bibinfo{volume}{366}},
  \bibinfo{pages}{426} (\bibinfo{year}{2002}).

\bibitem[{\citenamefont{Jankowski and Paldus}(1980)}]{jankowski:80}
\bibinfo{author}{\bibfnamefont{K.}~\bibnamefont{Jankowski}} \bibnamefont{and}
  \bibinfo{author}{\bibfnamefont{J.}~\bibnamefont{Paldus}},
  \bibinfo{journal}{Int. J. Quantum Chem.} \textbf{\bibinfo{volume}{18}},
  \bibinfo{pages}{1243} (\bibinfo{year}{1980}).

\bibitem[{\citenamefont{Kucharski and Bartlett}(1992)}]{kucharski:92}
\bibinfo{author}{\bibfnamefont{S.~A.} \bibnamefont{Kucharski}}
  \bibnamefont{and} \bibinfo{author}{\bibfnamefont{R.~J.}
  \bibnamefont{Bartlett}}, \bibinfo{journal}{J. Chem. Phys.}
  \textbf{\bibinfo{volume}{97}}, \bibinfo{pages}{4282} (\bibinfo{year}{1992}).

\bibitem[{\citenamefont{White}(1992)}]{white:92}
\bibinfo{author}{\bibfnamefont{S.~R.} \bibnamefont{White}},
  \bibinfo{journal}{Phys. Rev. Lett.} \textbf{\bibinfo{volume}{69}},
  \bibinfo{pages}{2863} (\bibinfo{year}{1992}).

\bibitem[{\citenamefont{Chan and Head-Gordon}(2002)}]{chan:02}
\bibinfo{author}{\bibfnamefont{G.~K.-L.} \bibnamefont{Chan}} \bibnamefont{and}
  \bibinfo{author}{\bibfnamefont{M.}~\bibnamefont{Head-Gordon}},
  \bibinfo{journal}{J. Chem. Phys.} \textbf{\bibinfo{volume}{116}},
  \bibinfo{pages}{4462} (\bibinfo{year}{2002}).

\bibitem[{\citenamefont{Liegener}(1985)}]{liegener:85}
\bibinfo{author}{\bibfnamefont{C.-M.} \bibnamefont{Liegener}},
  \bibinfo{journal}{J. Phys. C} \textbf{\bibinfo{volume}{18}},
  \bibinfo{pages}{6011} (\bibinfo{year}{1985}).

\bibitem[{\citenamefont{Joachim et~al.}(2000)\citenamefont{Joachim, Gimzewski,
  and Aviram}}]{joachim:00}
\bibinfo{author}{\bibfnamefont{C.}~\bibnamefont{Joachim}},
  \bibinfo{author}{\bibfnamefont{J.~K.} \bibnamefont{Gimzewski}},
  \bibnamefont{and} \bibinfo{author}{\bibfnamefont{A.}~\bibnamefont{Aviram}},
  \bibinfo{journal}{Nature} \textbf{\bibinfo{volume}{408}},
  \bibinfo{pages}{541} (\bibinfo{year}{2000}).

\bibitem[{\citenamefont{L{\"o}wdin}(1950)}]{lowdin:50}
\bibinfo{author}{\bibfnamefont{P.-O.} \bibnamefont{L{\"o}wdin}},
  \bibinfo{journal}{J. Chem. Phys.} \textbf{\bibinfo{volume}{18}},
  \bibinfo{pages}{365} (\bibinfo{year}{1950}).

\bibitem[{\citenamefont{Fisher-Hjalmars}(1965{\natexlab{a}})}]{hjalmars:65a}
\bibinfo{author}{\bibfnamefont{I.}~\bibnamefont{Fisher-Hjalmars}},
  \bibinfo{journal}{J. Chem. Phys.} \textbf{\bibinfo{volume}{42}},
  \bibinfo{pages}{1962} (\bibinfo{year}{1965}{\natexlab{a}}).

\bibitem[{\citenamefont{Fisher-Hjalmars}(1965{\natexlab{b}})}]{hjalmars:65b}
\bibinfo{author}{\bibfnamefont{I.}~\bibnamefont{Fisher-Hjalmars}},
  \bibinfo{journal}{Adv. Quantum Chem.} \textbf{\bibinfo{volume}{2}},
  \bibinfo{pages}{25} (\bibinfo{year}{1965}{\natexlab{b}}).

\bibitem[{\citenamefont{Parr}(1952)}]{parr:52}
\bibinfo{author}{\bibfnamefont{R.}~\bibnamefont{Parr}}, \bibinfo{journal}{J.
  Chem. Phys.} \textbf{\bibinfo{volume}{20}}, \bibinfo{pages}{1499}
  (\bibinfo{year}{1952}).

\bibitem[{\citenamefont{McWeeny}(1955)}]{mcweeny:55}
\bibinfo{author}{\bibfnamefont{R.}~\bibnamefont{McWeeny}},
  \bibinfo{journal}{Proc. R. Soc. London, Ser. A}
  \textbf{\bibinfo{volume}{227}}, \bibinfo{pages}{288} (\bibinfo{year}{1955}).

\bibitem[{\citenamefont{Parr}(1963)}]{parr:63}
\bibinfo{author}{\bibfnamefont{R.}~\bibnamefont{Parr}},
  \emph{\bibinfo{title}{The Quantum Theory of Molecular Electronic Structure}}
  (\bibinfo{publisher}{Benjamin}, \bibinfo{address}{New York},
  \bibinfo{year}{1963}).

\bibitem[{\citenamefont{Campbell et~al.}(1990)\citenamefont{Campbell, Gammel,
  and Loh}}]{campbell:90}
\bibinfo{author}{\bibfnamefont{D.~K.} \bibnamefont{Campbell}},
  \bibinfo{author}{\bibfnamefont{J.~T.} \bibnamefont{Gammel}},
  \bibnamefont{and} \bibinfo{author}{\bibfnamefont{E.~Y.} \bibnamefont{Loh},
  \bibfnamefont{Jr.}}, \bibinfo{journal}{Phys. Rev. B}
  \textbf{\bibinfo{volume}{42}}, \bibinfo{pages}{475} (\bibinfo{year}{1990}).

\bibitem[{\citenamefont{{Kouteck\'{y}}}(1967)}]{koutecky:67}
\bibinfo{author}{\bibfnamefont{J.}~\bibnamefont{{Kouteck\'{y}}}},
  \bibinfo{journal}{Chem. Phys. Lett.} \textbf{\bibinfo{volume}{1}},
  \bibinfo{pages}{249} (\bibinfo{year}{1967}).

\bibitem[{\citenamefont{{Del Re}}(1990)}]{delre:90}
\bibinfo{author}{\bibfnamefont{G.}~\bibnamefont{{Del Re}}},
  \bibinfo{journal}{Int. J. Quantum Chem.} \textbf{\bibinfo{volume}{37}},
  \bibinfo{pages}{479} (\bibinfo{year}{1990}).

\bibitem[{\citenamefont{Mataga and Nishimoto}(1957)}]{mataga:57}
\bibinfo{author}{\bibfnamefont{N.}~\bibnamefont{Mataga}} \bibnamefont{and}
  \bibinfo{author}{\bibfnamefont{K.}~\bibnamefont{Nishimoto}},
  \bibinfo{journal}{Z. Phys. Chem. (Frankfurt am Main)}
  \textbf{\bibinfo{volume}{13}}, \bibinfo{pages}{140} (\bibinfo{year}{1957}).

\bibitem[{\citenamefont{Ohno}(1964)}]{ohno:64}
\bibinfo{author}{\bibfnamefont{K.}~\bibnamefont{Ohno}},
  \bibinfo{journal}{Theor. Chim. Acta} \textbf{\bibinfo{volume}{2}},
  \bibinfo{pages}{219} (\bibinfo{year}{1964}).

\bibitem[{\citenamefont{{Kouteck\'{y}}
  et~al.}(1985)\citenamefont{{Kouteck\'{y}}, Paldus, and {{\v C}{\'
  \i}{\v{z}ek}}}}]{koutecky:85}
\bibinfo{author}{\bibfnamefont{J.}~\bibnamefont{{Kouteck\'{y}}}},
  \bibinfo{author}{\bibfnamefont{J.}~\bibnamefont{Paldus}}, \bibnamefont{and}
  \bibinfo{author}{\bibfnamefont{J.}~\bibnamefont{{{\v C}{\' \i}{\v{z}ek}}}},
  \bibinfo{journal}{J. Chem. Phys.} \textbf{\bibinfo{volume}{83}},
  \bibinfo{pages}{1722} (\bibinfo{year}{1985}).

\bibitem[{\citenamefont{H{\"u}ckel}(1931)}]{huckel:31}
\bibinfo{author}{\bibfnamefont{E.}~\bibnamefont{H{\"u}ckel}},
  \bibinfo{journal}{Z. Phys.} \textbf{\bibinfo{volume}{70}},
  \bibinfo{pages}{204} (\bibinfo{year}{1931}).

\bibitem[{\citenamefont{Paldus et~al.}(1982)\citenamefont{Paldus, {\v C}{\'
  \i}{\v z}ek, and Huba{\v c}}}]{paldus:74}
\bibinfo{author}{\bibfnamefont{J.}~\bibnamefont{Paldus}},
  \bibinfo{author}{\bibfnamefont{J.}~\bibnamefont{{\v C}{\' \i}{\v z}ek}},
  \bibnamefont{and} \bibinfo{author}{\bibnamefont{Huba{\v c}}},
  \bibinfo{journal}{Int. J. Quantum Chem.} \textbf{\bibinfo{volume}{22}},
  \bibinfo{pages}{1281} (\bibinfo{year}{1982}).

\bibitem[{\citenamefont{Pariser}(1956)}]{pariser:56}
\bibinfo{author}{\bibfnamefont{R.}~\bibnamefont{Pariser}}, \bibinfo{journal}{J.
  Chem. Phys.} \textbf{\bibinfo{volume}{24}}, \bibinfo{pages}{250}
  (\bibinfo{year}{1956}).

\bibitem[{\citenamefont{Coulson and Rushbrooke}(1940)}]{coulson:40}
\bibinfo{author}{\bibfnamefont{C.~A.} \bibnamefont{Coulson}} \bibnamefont{and}
  \bibinfo{author}{\bibfnamefont{G.~S.} \bibnamefont{Rushbrooke}},
  \bibinfo{journal}{Proc. Camb. Philos. Soc.} \textbf{\bibinfo{volume}{36}},
  \bibinfo{pages}{193} (\bibinfo{year}{1940}).

\bibitem[{\citenamefont{Koopmans}(1934)}]{koopmans:34}
\bibinfo{author}{\bibfnamefont{T.}~\bibnamefont{Koopmans}},
  \bibinfo{journal}{Physica (Utrecht)} \textbf{\bibinfo{volume}{1}},
  \bibinfo{pages}{105} (\bibinfo{year}{1934}).

\bibitem[{\citenamefont{Abramowitz and Stegun}(1968)}]{abramowitz:68}
\bibinfo{editor}{\bibfnamefont{M.}~\bibnamefont{Abramowitz}} \bibnamefont{and}
  \bibinfo{editor}{\bibfnamefont{I.}~\bibnamefont{Stegun}}, eds.,
  \emph{\bibinfo{title}{Handbook of Mathematical Functions}}
  (\bibinfo{publisher}{U.S. GPO}, \bibinfo{address}{{Washington, D.C.}},
  \bibinfo{year}{1968}), \bibinfo{note}{p. 811}.

\bibitem[{\citenamefont{Monkhorst et~al.}(1981)\citenamefont{Monkhorst,
  Jeziorski, and Harris}}]{monkhorst:81}
\bibinfo{author}{\bibfnamefont{H.~J.} \bibnamefont{Monkhorst}},
  \bibinfo{author}{\bibfnamefont{B.}~\bibnamefont{Jeziorski}},
  \bibnamefont{and} \bibinfo{author}{\bibfnamefont{F.~E.}
  \bibnamefont{Harris}}, \bibinfo{journal}{Phys. Rev. A}
  \textbf{\bibinfo{volume}{23}}, \bibinfo{pages}{1639} (\bibinfo{year}{1981}).

\bibitem[{\citenamefont{Stolarczyk and Monkhorst}(1984)}]{stolarczyk:84}
\bibinfo{author}{\bibfnamefont{L.~Z.} \bibnamefont{Stolarczyk}}
  \bibnamefont{and} \bibinfo{author}{\bibfnamefont{H.~J.}
  \bibnamefont{Monkhorst}}, \bibinfo{journal}{Int. J. Quantum Chem. (Quantum
  Chem. Symposium)} \textbf{\bibinfo{volume}{18}}, \bibinfo{pages}{267}
  (\bibinfo{year}{1984}).

\bibitem[{\citenamefont{Brandow}(1967)}]{brandow:67}
\bibinfo{author}{\bibfnamefont{B.~H.} \bibnamefont{Brandow}},
  \bibinfo{journal}{Rev. Mod. Phys.} \textbf{\bibinfo{volume}{39}},
  \bibinfo{pages}{771} (\bibinfo{year}{1967}).

\bibitem[{\citenamefont{Beleznay et~al.}(1981)\citenamefont{Beleznay, Suhai,
  and Ladik}}]{beleznay:81}
\bibinfo{author}{\bibfnamefont{F.}~\bibnamefont{Beleznay}},
  \bibinfo{author}{\bibfnamefont{S.}~\bibnamefont{Suhai}}, \bibnamefont{and}
  \bibinfo{author}{\bibfnamefont{J.}~\bibnamefont{Ladik}},
  \bibinfo{journal}{Int. J. Quantum Chem.} \textbf{\bibinfo{volume}{20}},
  \bibinfo{pages}{683} (\bibinfo{year}{1981}).

\bibitem[{\citenamefont{Press et~al.}(1992)\citenamefont{Press, Teukolsky,
  Vetterling, and Flannery}}]{numrecC:92}
\bibinfo{author}{\bibfnamefont{W.~H.} \bibnamefont{Press}},
  \bibinfo{author}{\bibfnamefont{S.~A.} \bibnamefont{Teukolsky}},
  \bibinfo{author}{\bibfnamefont{W.~T.} \bibnamefont{Vetterling}},
  \bibnamefont{and} \bibinfo{author}{\bibfnamefont{B.~P.}
  \bibnamefont{Flannery}}, \emph{\bibinfo{title}{Numerical Recipes in C}}
  (\bibinfo{publisher}{CUP}, \bibinfo{address}{Cambridge},
  \bibinfo{year}{1992}), \bibinfo{note}{chapter 4}.

\bibitem[{\citenamefont{Metzner and Vollhardt}(1989)}]{metzner:89}
\bibinfo{author}{\bibfnamefont{W.}~\bibnamefont{Metzner}} \bibnamefont{and}
  \bibinfo{author}{\bibfnamefont{D.}~\bibnamefont{Vollhardt}},
  \bibinfo{journal}{Phys. Rev. B} \textbf{\bibinfo{volume}{39}},
  \bibinfo{pages}{4462} (\bibinfo{year}{1989}).

\bibitem[{\citenamefont{Toyozawa}(1954)}]{toyozawa:54}
\bibinfo{author}{\bibfnamefont{Y.}~\bibnamefont{Toyozawa}},
  \bibinfo{journal}{Prog. Theor. Phys.} \textbf{\bibinfo{volume}{12}},
  \bibinfo{pages}{421} (\bibinfo{year}{1954}).

\bibitem[{\citenamefont{Kunz}(1972)}]{kunz:72}
\bibinfo{author}{\bibfnamefont{A.~B.} \bibnamefont{Kunz}},
  \bibinfo{journal}{Phys. Rev. B} \textbf{\bibinfo{volume}{6}},
  \bibinfo{pages}{606} (\bibinfo{year}{1972}).

\bibitem[{\citenamefont{Pantelides et~al.}(1974)\citenamefont{Pantelides,
  Mickish, and Kunz}}]{pantelides:74}
\bibinfo{author}{\bibfnamefont{S.~T.} \bibnamefont{Pantelides}},
  \bibinfo{author}{\bibfnamefont{D.~J.} \bibnamefont{Mickish}},
  \bibnamefont{and} \bibinfo{author}{\bibfnamefont{A.~B.} \bibnamefont{Kunz}},
  \bibinfo{journal}{Phys. Rev. B} \textbf{\bibinfo{volume}{10}},
  \bibinfo{pages}{2602} (\bibinfo{year}{1974}).

\bibitem[{\citenamefont{Mila and Penc}(2001)}]{mila:01}
\bibinfo{author}{\bibfnamefont{F.}~\bibnamefont{Mila}} \bibnamefont{and}
  \bibinfo{author}{\bibfnamefont{K.}~\bibnamefont{Penc}}, \bibinfo{journal}{J.
  Electron Spectrosc. Relat. Phenom.} \textbf{\bibinfo{volume}{117-118}},
  \bibinfo{pages}{451} (\bibinfo{year}{2001}).

\bibitem[{\citenamefont{Aissing and Monkhorst}(1993)}]{aissing:93}
\bibinfo{author}{\bibfnamefont{G.}~\bibnamefont{Aissing}} \bibnamefont{and}
  \bibinfo{author}{\bibfnamefont{H.~J.} \bibnamefont{Monkhorst}},
  \bibinfo{journal}{Int. J. Quantum Chem. (Quantum Chem. Symposium)}
  \textbf{\bibinfo{volume}{27}}, \bibinfo{pages}{81} (\bibinfo{year}{1993}).

\bibitem[{\citenamefont{Mukherjee and Pal}(1989)}]{mukherjee:89}
\bibinfo{author}{\bibfnamefont{D.}~\bibnamefont{Mukherjee}} \bibnamefont{and}
  \bibinfo{author}{\bibfnamefont{S.}~\bibnamefont{Pal}}, \bibinfo{journal}{Adv.
  Quantum Chem.} \textbf{\bibinfo{volume}{20}}, \bibinfo{pages}{291}
  (\bibinfo{year}{1989}).

\end{thebibliography}

\end{document}